\documentclass[prd,preprint,tightenlines,floatfix,showpacs,preprintnumbers,nofootinbib,eqsecnum]{revtex4}

 \usepackage[dvips,final]{graphicx}
  \usepackage{amssymb}
   \usepackage{amsmath}
    \usepackage{amsfonts}
     \usepackage{epsfig}
     \usepackage{color}
      \usepackage{bm} 

\usepackage{booktabs}
\usepackage{multirow}
\usepackage{slashed}
\usepackage{comment}
 \newcommand\la{\langle}
 \newcommand\ra{\rangle}
 \newcommand\beq{\begin{equation}}
 
 \newcommand\eeq{\end{equation}}
 \newcommand\beqn{\begin{eqnarray}}
 \newcommand\eeqn{\end{eqnarray}}

 \def\gsim{\mathrel{\rlap{\lower4pt\hbox{\hskip1pt$\sim$}}
 \raise1pt\hbox{$>$}}}

 \def\GeV{\,\mbox{GeV}}
 
 \def\lsim{\mathrel{\rlap{\lower4pt\hbox{\hskip1pt$\sim$}}
     \raise1pt\hbox{$<$}}}         
 \def\gsim{\mathrel{\rlap{\lower4pt\hbox{\hskip1pt$\sim$}}
     \raise1pt\hbox{$>$}}}         
 \def\Re{\,\mbox{Re}\,}
 \def\Im{\,\mbox{Im}\,}

 \def\beq{\begin{equation}}
 \def\eeq{\end{equation}}
 
 \def\beqy{\begin{eqnarray}}
 \def\eeqy{\end{eqnarray}}

 \def\sqqbar{\sigma_{q \bar q}^N}
 
 \def\a{\alpha}
 \def\xBj{x_{Bj}}

\begin{document}


\title
{
Nuclear shadowing in DIS at electron-ion colliders
}

\author{Michal Krelina$^{1,2}$}
\email{michal.krelina@fjfi.cvut.cz}

\author{Jan Nemchik$^{2,3}$}
\email{nemcik@saske.sk}

\affiliation{
{$^1$\sl
Departamento de F\'{\i}sica,
Universidad T\'ecnica Federico Santa Mar\'{\i}a,
Casilla 110-V, Valpara\'{\i}so, Chile
}\vspace{0.5cm}\\
{$^2$\sl
Czech Technical University in Prague, FNSPE, B\v rehov\'a 7, 11519
Prague, Czech Republic
}\vspace{0.5cm}\\
{$^3$\sl
Institute of Experimental Physics SAS, Watsonova 47, 04001 Ko\v
sice, Slovakia
}
\vspace*{1.0cm}}
\begin{abstract}
\vspace*{1.0cm}
We present a revision of predictions for nuclear shadowing in deep-inelastic scattering at small Bjorken $\xBj$ corresponding to kinematic regions accessible by the future experiments at electron-ion colliders. The nuclear shadowing is treated within the color dipole formalism based on the rigorous  Green function technique. This allows incorporating naturally color transparency and coherence length effects, which are not consistently and properly included in present calculations. For the lowest $|q\bar q\ra$ Fock component of the photon, our calculations
are based on an exact numerical solution of the evolution equation for the Green function. Here the magnitude of shadowing is tested using a realistic form for the nuclear density function, as well as various phenomenological models for the dipole cross section. The corresponding variation of the transverse size of the $q\bar q$ photon fluctuations is important for $\xBj\gsim 10^{-4}$, on the contrary with the most of other models, which use frequently only the eikonal approximation with the ``frozen" transverse size.
At $\xBj\lsim 0.01$ we calculate within the same formalism also a shadowing correction for the higher Fock component of the photon containing gluons. The corresponding magnitudes of gluon shadowing correction are compared adopting different phenomenological dipole models. Our results are tested by available data from the E665 and NMC collaborations. Finally, the magnitude of nuclear shadowing is predicted for various kinematic regions that should be scanned by the future experiments at electron-ion colliders.
\end{abstract}

\pacs{24.85.+p,13.60.Le,13.85.Lg}

\maketitle

%
%
%
\section{Introduction}
\label{Sec:Intro}
%
%
%

One of the main topic, which is proposed to be studied by the future experiments at electron-ion colliders (EICs) \cite{Accardi:2012qut,Aschenauer:2014cki,Aschenauer:2017jsk,AbelleiraFernandez:2012cc} using present RHIC and LHC facilities, represents the nuclear shadowing at small values of Bjorken $\xBj$. This gives the main motivation for study of possible sources causing a suppression not only in deep-inelastic scattering (DIS) off nuclei (see \cite{Nemchik:2003wx,Kopeliovich:2008ek}, for example), but also in other exclusive processes at large energies, like diffractive photo- and electroproduction of vector mesons on nuclei (see \cite{Kopeliovich:1991pu,Kopeliovich:1993gk,Kopeliovich:1993pw,Kopeliovich:2001xj,Kopeliovich:2007wx}, for example), the Drell-Yan process (see \cite{Kopeliovich:2001hf,Goncalves:2016qku,Basso:2016ulb}, for example), as well as the inclusive hadron production in proton-nucleus collisions (see \cite{Kopeliovich:2002yh,Nemchik:2008xy,cronin-13}, for example).

The convenient, frequently used and experimentally measured observable for study of nuclear shadowing in DIS is the nuclear structure function $F_2^A$. In the region of small $\xBj$ the effect of nuclear shadowing manifests itself as an inequality $F_2^A/(A~F_2^N) < 1$, where $A$ is the number of nucleons in a nuclear target and $F_2^N$ represents the free nucleon structure function (see the review \cite{Arneodo:1992wf}, for example). Such a study of shadowing can contribute to our understanding and interpretation of suppression observed by hadron-nucleus and heavy-ion experiments as well as allows to predict the corresponding nuclear phenomena expected by the future measurements at EICs.

Interpretation of nuclear shadowing depends on the reference frame. In the present paper, the shadowing is treated within the light-cone (LC) color dipole formalism, which describes this phenomenon in the rest frame of the nucleus as the nuclear shadowing of hadronic components of the virtual photon caused by their multiple scattering inside the target \cite{Bauer:1977iq,Frankfurt:1988nt,Brodsky:1989qz,Brodsky:2004qa,Nikolaev:1990ja,Melnitchouk:1993vc,Nikolaev:1994de,Piller:1995kh,Kopeliovich:1995yr,Kopeliovich:1995ju,Piller:1999wx,Kopeliovich:2000ra,Nemchik:2003wx,Kopeliovich:2008ek}. However, the interpretation in the infinite momentum frame treats the same phenomenon as a result of parton fusion  \cite{Kancheli:1973vc,Gribov:1984tu,Mueller:1985wy,Qiu:1986wh},
leading to a reduction of the parton density at low Bjorken $\xBj$.

The destructive interference of amplitudes corresponding to interactions, which occur on different bound nucleons, leads to the effect known as the \textit{quantum coherence}. It controls the dynamics of nuclear shadowing and can be interpreted also as the lifetime $t_c$ of the photon fluctuations. Assuming, for example, the lowest $|q\bar q\ra$ Fock component of the photon, this lifetime can be expressed relying on the uncertainty principle and Lorentz time dilation and has the following form,
%
\beq
t_c = \frac{2\,\nu}{Q^2 + M_{q\bar q}^2}\ ,
\label{10}
 \eeq
%
where $\nu$ and $Q^2$ is the photon energy and virtuality, and $M_{\bar qq}$ is the effective mass of the $q\bar q$ pair. In what follows, besides the term \textit{coherence time} $t_c$, we will use also the term \textit{coherence length} (CL) $l_c$, since the light-cone kinematics is assumed, $l_c=t_c$. The CL is related to the longitudinal momentum transfer as $q_c=1/l_c$.

In the most of present calculations, the effect of quantum coherence is not included properly and the nuclear shadowing is calculated relying on eikonal approximation, which is effective only at sufficiently large $l_c\gg R_A$, where $R_A$ is the nuclear radius. The magnitude of the coherence length given by Eq.~(\ref{10}) for the case of the $|q\bar q\ra$ state is larger than for higher Fock states containing gluons, $|q\bar qG\ra, |q\bar qGG\ra, ...$, since the corresponding effective masses are larger than $M_{q\bar q}$. For this reason, the shadowing caused by gluons manifests itself at much higher energies (much smaller $\xBj$) than that caused by quarks. Consequently, in the present paper we will treat the nuclear shadowing with no restrictions on a magnitude of the coherence length using the Green function formalism
\cite{Kopeliovich:1991pu,Kopeliovich:2001xj,Kopeliovich:1999am,Nemchik:2003wx,Kopeliovich:2007wx,Kopeliovich:2008ek,Goncalves:2016qku}.

The important ingredient of the color dipole approach is the dipole cross section, $\sigma_{q\bar q}(r)$, which represents the interaction of $q\bar q$ dipole of transverse separation $\vec r$ with a nucleon \cite{Kopeliovich:1981pz}. The flavor invariance due to universality of the QCD coupling, and the small size behavior, $\sigma_{q\bar q}(r) \propto r^2$ for $r\!\!\to\!0$ (the property known as the \textit{color transparency}) are
two main properties of $\sigma_{q\bar q}(r)$ leading to its energy- and $r$-universality as well as to its potentiality to describe in an uniform way various high-energy processes.

The magnitude of nuclear shadowing is strongly correlated with the shape of $\sigma_{q\bar q}(r)$, which cannot be predicted from the first principles because of poorly known higher order perturbative QCD (pQCD) corrections and nonperturbative effects. However, it can be separately determined from the $ep$ DIS data at HERA, what allows to include naturally in our calculations all higher order corrections and higher twist effects. For this reason, we are forced to rely on a number of phenomenological parametrizations proposed in the literature (see \cite{GolecBiernat:1998js,Kowalski:2006hc,Kopeliovich:1999am,Bartels:2002cj,Rezaeian:2012ji}, for example) which are based on the fits to the HERA DIS data.

Consequently, in the current paper we present for the first time a detailed analysis of shadowing effects in DIS off nuclei within the color dipole formalism revising and improving the old calculations from \cite{Nemchik:2003wx,Kopeliovich:2008ek}. The main motivation for such a study is based on the new data expected from the planned realization of future measurements at EICs. Consequently, our predictions for shadowing will be predominantly focused on corresponding kinematic regions.
The magnitude of shadowing coming from the lowest $|q\bar q\ra$ (quark shadowing) as well as from higher $|q\bar qG\ra, \dots$ (gluon shadowing) Fock components of the photon is calculated using the Green function technique, which naturally includes not only the effects of quantum coherence but also color transparency. Besides, we extend for the first time such a study also to various phenomenological dipole models for $\sigma_{q\bar q}(r)$. Consequently, the corresponding variations in model predictions using distinct phenomenological parametrizations of $\sigma_{q\bar q}(r)$ can be treated as a measure of the underlined theoretical uncertainty. This has a large impact for future studies of the QCD dynamics, mainly in connection with determination of the onset of shadowing effects not only in DIS off nuclei, but also in other processes occurring in lepton (proton)-nucleus interactions and in heavy-ion collisions.

Another additional innovative output of the present paper in based on our provision of
public numerical grids for the magnitude of gluon shadowing, $R_G$, calculated as the shadowing correction from the $|q\bar qG\ra$ Fock component of the photon, for various nuclear targets, photon virtualities $Q^2$, values of the Bjorken $x_{Bj}$, as well as for various values of the nuclear impact parameter $b$. The corresponding numerical values can be found on the following Zenodo web-page: \url{https://zenodo.org/record/3470138} \cite{dataset}.

The paper is organized as follows. In the next Section~\ref{lc} we present a short review of the light-cone dipole phenomenology for description of nuclear shadowing in DIS which is based on the Green function formalism. Here, we treat separately the shadowing correction from the lowest $|q\bar q\ra$ component of the photon as well as from the higher Fock states containing gluons.
Consequently, in Sect.~\ref{glue-all} we discuss contributions to gluon shadowing from different multi-gluon Fock components with respect to kinematic regions accessible by the future experiments at EICs. Here we show an importance of the onset of quantum coherence effects which are not treated properly in present calculations.
The next Section~\ref{results}  is devoted to model predictions for the shadowing and to comparison of the corresponding numerical results with available data. Here, we present our results
including various photon fluctuations, such as  $q\bar q$ and $q\bar qG$, as function of Bjorken $\xBj$ and photon virtuality $Q^2$, especially for kinematic regions scanned by the future measurements at EICs.
Here, we provide for the first time also model predictions adopting various phenomenological models of the dipole cross sections and test their impact on the magnitude of shadowing.
We also compare our results with the onset of shadowing effects obtained from the parton model within a broad range of $\xBj$.
The possibility to obtain directly numerical values for the gluon shadowing factor from the public grid is mentioned in Sect.~\ref{datasets}.
Finally, in Section~\ref{conclusions} we summarize our main results and analyze theoretical uncertainties in estimation of nuclear shadowing in DIS as a function of $\xBj$ and $Q^2$. Here, we also discuss the possibility of an experimental evidence of a gluon contribution to the overall nuclear shadowing in kinematic regions scanned by the future experiments at EICs. 

%
%
%
\section{A short review of the light-cone dipole approach to nuclear shadowing}
\label{lc}
%
%
%

In the rest frame of the nucleus, different Fock components of the virtual photon can contribute to nuclear shadowing effect in the total virtual photoabsorption cross section $\sigma_{tot}^{\gamma^*A}(\xBj,Q^2)$ or in the structure function $F_2^A(\xBj,Q^2)$. Consequently, the nuclear cross section $\sigma_{tot}^{\gamma^*A}(\xBj,Q^2)$ then reads,
%
 \beq
\sigma_{tot}^{\gamma^*A}(\xBj,Q^2) =
A~\sigma_{tot}^{\gamma^*N}(\xBj,Q^2) -
\Delta\sigma_{tot}(\xBj,Q^2)\, ,
\label{110}
 \eeq
%
where $\Delta\sigma_{tot}(\xBj,Q^2)$ represents the summation of shadowing corrections coming from different Fock states, $|q\bar q\ra, |q\bar qG\ra, |q\bar q2G\ra, \cdots$
%
 \beq
\Delta\sigma_{tot}(\xBj,Q^2) =
\Delta\sigma_{tot}^{q\bar q}(\xBj,Q^2) +
\Delta\sigma_{tot}^{q\bar qG}(\xBj,Q^2) +
\Delta\sigma_{tot}^{q\bar q2G}(\xBj,Q^2) + \cdots
\, .
\label{112}
 \eeq
%

In Eq.~(\ref{110}) the variable $\sigma_{tot}^{\gamma^*N}(\xBj,Q^2)$ represents the total virtual photoabsorption cross section on a nucleon defined at Bjorken variable $\xBj$ given by
%
\beq
\xBj = \frac{Q^2}{2\,m_N\,\nu} \approx
\frac{Q^2}{Q^2 + s}\, ,
\label{115}
\eeq
%
where $s$ is the $\gamma^*$-nucleon center of mass (c.m.) energy
squared and $m_N$ is the mass of the nucleon.

In the light-cone dipole approach, the variable $\sigma_{tot}^{\gamma^*N}(\xBj,Q^2)$ can be expressed in the quantum-mechanical form,
%
 \beq
\sigma_{tot}^{\gamma^*N}(x_{Bj},Q^2) =
\la\Psi_{q\bar q}(r,\alpha)|\sigma_{q\bar q}(r)|\Psi_{q\bar q}(r,\alpha)\ra
=
\int d^2 r \int_{0}^{1} d\alpha\,\Bigl
| \Psi_{q\bar q}(\vec{r},\alpha,Q^2)\,\Bigr |^2
~\sigma_{q\bar q}(\vec{r},s)\, ,
\label{120}
 \eeq
%
where $\Psi_{q\bar q}({\vec{r}},\alpha,Q^2)$ is the LC wave function of the $|q\bar q\ra$ Fock component of the photon, which depends also on the photon virtuality $Q^2$ and the relative share $\alpha$ of the photon momentum carried by the quark and $\sigma_{q\bar q}({\vec{r}},s)$ is the dipole cross section, which depends on the $q\bar q$ transverse
separation $\vec{r}$ and the c.m. energy squared $s$. Here, the corresponding dependence on $\xBj$ is related to $s$ via Eq.~(\ref{115}).

The dipole cross section $\sigma_{q\bar q}(r,\xBj)$, representing the essential ingredient of the color dipole approach, has been first introduced in Ref.~\cite{Kopeliovich:1981pz}.
Two main properties of $\sigma_{q\bar q}(r,\xBj)$, such as the flavor invariance due to universality of the QCD coupling, and the small size behavior, $\sigma_{q\bar q}(r)\propto r^2$ for $r\to 0$ (the property known as the \textit{color transparency} \cite{Kopeliovich:1981pz,Bertsch:1981py,Brodsky:1988xz}), support its flavor independence as well as the energy and transverse size universality.

Because of poorly known higher order perturbative QCD corrections and nonperturbative effects, the cross section $\sigma_{q\bar q}(r,\xBj)$ cannot be predicted reliably. Here, we are forced to use phenomenological parametrizations of $\sigma_{q\bar q}(r,\xBj)$ based on
fits to HERA data on DIS and structure functions. Although about ten different parametrizations can be found recently in the literature, for our study of the onset of nuclear shadowing effects we use the most popular of them denoted as GBW \cite{GolecBiernat:1998js,Kowalski:2006hc}, KST \cite{Kopeliovich:1999am}, BGBK \cite{Bartels:2002cj} and IP-Sat \cite{Rezaeian:2012ji}.

Another ingredient, which is important in calculations of nuclear shadowing, is the LC
perturbative distribution amplitude (``wave function'') of the photon. For the lowest
$|q\bar q\ra$ Fock component of the photon, it has the following form for transversally (T) and longitudinally (L) polarized photons \cite{Kogut:1969xa,Bjorken:1970ah,Nikolaev:1990ja}:
%
 \beq
\Psi_{q\bar q}^{T,L}({\vec{r}},\alpha,Q^2) =
\frac{\sqrt{N_{C}\,\alpha_{em}}}{2\,\pi}\,\,
Z_{q}\,\bar{\chi}\,\hat{O}^{T,L}\,\chi\,
K_{0}(\epsilon\,r),
\label{122}
 \eeq
%
where $\chi$ and $\bar{\chi}$ are the spinors of the quark and
antiquark respectively, $Z_{q}$ is the quark charge, $N_{C} = 3$ is
the number of colors, and $K_{0}(\epsilon r)$ is a modified Bessel
function with
%
 \beq
\epsilon^{2} =
\alpha\,(1-\alpha)\,Q^{2} + m_{q}^{2}\ ,
\label{123}
 \eeq
%
where $m_{q}$ is the quark mass.

The energy dependence of the hadron production cross section can be included in two different alternative ways. Relying on the two-gluon approximation \cite{Kopeliovich:1981pz}, the dipole cross section is constant and energy dependence comes from the higher-order corrections related to gluon radiation. Another way is to involve higher Fock components of the photon containing gluons in addition to the lowest $|q\bar q\ra$ state. Here, we prefer the former way introducing the energy (Bjorken $\xBj$) dependence in the dipole cross section $\sigma_{q\bar q}(r,s)$ not including higher Fock states into the photon wave function
as is expressed by Eq.~(\ref{120}).

The operators $\widehat{O}^{T,L}$ in Eq.~(\ref{122}) read,
%
 \beqn
\widehat{O}^{T} = m_{q}\,\,\vec{\sigma}\cdot\vec{e} +
i\,(1-2\alpha)\,(\vec{\sigma}\cdot\vec{n})\,
(\vec{e}\cdot\vec{\nabla}_r) + (\vec{\sigma}\times
\vec{e})\cdot\vec{\nabla}_r\,,
\hspace*{0.8cm}
\widehat{O}^{L} =
2\,Q\,\alpha (1 - \alpha)\,(\vec{\sigma}\cdot\vec{n})\,,
\nonumber\\
 \label{124}
 \eeqn
%
where $\vec\nabla_r$ acts on the transverse coordinate $\vec r$,
$\vec{e}$ is the polarization vector of the photon, $\vec{n}$ is a
unit vector parallel to the photon momentum, and $\vec{\sigma}$ is
the three vector of the Pauli spin-matrices.

The transverse size of the $q\bar q$ photon fluctuation is controlled by the distribution amplitude Eq.~(\ref{122}) with the corresponding mean value
%
 \beq
\la r\ra \sim \frac{1}{\epsilon} =
\frac{1}{\sqrt{Q^{2}\,\alpha\,(1-\alpha) + m_{q}^{2}}}\,.
\label{130}
 \eeq
%
Within the pQCD, very asymmetric $q\bar q$ pairs with $\alpha$ or $(1-\alpha) \lsim
m_q^2/Q^2$ lead to a huge magnitude of the mean transverse separation $\la r\ra \sim 1/m_q$
due to small current quark masses. In order to solve this problem, we are forced to rely on
a popular recipe introducing an effective quark mass $m_{eff}\sim \Lambda_{QCD}$, which represents the nonperturbative interaction effects between the $q$ and $\bar q$. However, here we prefer another more consistent and straightforward way using a phenomenology based on the light-cone Green function formalism \cite{Kopeliovich:1999am} where such $q-\bar q$ interquark interaction is explicitly included.

The propagation of an interacting $q\bar q$ pair between points with longitudinal coordinates $z_1$ and $z_2$ and with initial and final transverse separations $\vec{r_1}$ and $\vec{r_2}$ is described by the Green function $G_{q\bar q}(\vec{r_2},z_2;\vec{r_1},z_1)$ satisfying the following two-dimensional Schr\"odinger equation,
%
 \beq
i\frac{d}{dz_2}\,G_{q\bar q}(\vec{r_2},z_2;\vec{r_1},z_1)=
\left[\frac{\epsilon^{2} - \Delta_{r_{2}}}{2\,\nu\,\alpha\,(1-\alpha)}
+V_{q\bar q}(z_2,\vec{r_2},\alpha)\right]
G_{q\bar q}(\vec{r_2},z_2;\vec{r_1},z_1)\ ,
\label{135}
 \eeq
%
with the boundary condition
%
 \beq
G_{q\bar q}(\vec{r_2},z_2;\vec{r_1},z_1)|_{z_2=z_1}=
\delta^2(\vec{r_1}-\vec{r_2})\, .
\label{136}
 \eeq
%

Considering the propagation of a $q\bar q$ pair in vacuum, the LC potential $V_{q\bar q}(z_2,\vec{r_2},\alpha)$ in (\ref{135}) contains only the real part, which describes the
interaction between the $q$ and $\bar{q}$. Although, more realistic models for $\Re V_{q\bar q}$ can be found in the literature (see \cite{Pirner:2004qd,Pirner:2009zz}, for example)
for the sake of simplicity we use an oscillator form of this potential,
%
 \beq
{\Re}\,V_{q\bar q}(z_2,\vec{r_2},\alpha) =
\frac{a^4(\alpha)\,\vec{r_2}\,^2}
{2\,\nu\,\alpha(1-\alpha)}\ ,
\label{140}
 \eeq
%
what leads to an analytic solution of the corresponding Schr\"odinger equation (\ref{135}) for the light-cone Green function, which has the following form,
%
 \beqn
G_{q\bar q}(\vec{r_2},z_2;\vec{r_1},z_1)
&=&
\frac{a^2(\alpha)}{2\;\pi\;i\;
{\rm sin}(\omega\,\Delta z)}\, {\rm exp}
\left\{\frac{i\,a^2(\alpha)}{{\rm sin}(\omega\,\Delta z)}\,
\Bigl[(r_1^2+r_2^2)\,{\rm cos}(\omega \;\Delta z) -
2\;\vec{r_1}\cdot\vec{r_2}\Bigr]\right\}
\nonumber\\
&&
\times {\rm exp}\left[-
\frac{i\,\epsilon^{2}\,\Delta z}
{2\,\nu\,\alpha\,(1-\alpha)}\right] \ ,
\label{142}
 \eeqn
%
where $\Delta z=z_2-z_1$, and $\omega = \frac{a^2(\alpha)}{\nu\;\alpha(1-\alpha)}$.
The shape of the function $a(\alpha)$ in Eqs.~(\ref{140}) and (\ref{142}) has been determined in Ref.~\cite{Kopeliovich:1999am} with parameters adjusted to the data on the total photoabsorption cross section, diffractive proton dissociation and shadowing in nuclear photoabsorption reaction.
Here, we would like to emphasize that any form of the $q-\bar q$ interaction potential should be consistent with the mean $q-\bar q$ and quark-gluon transverse separations, which matter for shadowing and are determined from the fit to diffraction data. We checked that our choice of the potential given by Eq.~(\ref{140}) complies with this condition.

From the known shape of the LC Green function, one can calculate the probability amplitude to find the $q\bar q$ fluctuation of a photon at the longitudinal coordinate $z_2$ with a transverse separation $\vec r$ as the integral over the point $z_1$ where the $q\bar q$ pair is created by the photon with initial separation zero,
%
 \beq
\Psi^{T,L}_{q\bar q}(\vec r,\alpha)=
\frac{i\,Z_q\sqrt{\alpha_{em}}}
{4\pi\,E\,\alpha(1-\alpha)}
\int\limits_{-\infty}^{z_2}dz_1\,
\Bigl(\bar\chi\;\widehat O^{T,L}\chi\Bigr)\,\,
G_{q\bar q}(\vec{r},z_2;\vec{r_1},z_1)
\Bigr|_{r_1=0}\ .
\label{146}
 \eeq
%
Here, the operators $\widehat O^{T,L}$ are defined by Eq.~(\ref{124}) and act on the coordinate $\vec r_1$.

Using the following expression for the transverse part,
%
 \beq
\bar\chi\;\widehat O^{T}\chi
=
\bar\chi\;m_{q}\,\,\vec{\sigma}\cdot\vec{e}\,\chi +
\bar\chi\;\left[i\,(1-2\alpha)\,(\vec{\sigma}\cdot\vec{n})\,
\vec{e} + (\vec{\sigma}\times
\vec{e})\right]\,\chi\cdot\vec{\nabla}_{r}=
E+\vec F\cdot\vec\nabla_{r}\ ,
\label{150}
 \eeq
%
 then the distribution functions for the $q\bar q$ fluctuation of the photon, accounting for the $q-\bar q$ interaction, read
%
 \beq
\Psi^{T}_{q\bar q}(\vec r,\alpha) =
Z_q\sqrt{\alpha_{em}}\,\left[E\,\Phi_0(\epsilon,r,\lambda)
+ \vec F\,\vec\Phi_1(\epsilon,r,\lambda)\right]\ ,
\label{152}
 \eeq
%
%
 \beq
\Psi^{L}_{q\bar q}(\vec r,\alpha) =
2\,Z_q\sqrt{\alpha_{em}}\,Q\,\alpha(1-\alpha)\,
\bar\chi\;\vec\sigma\cdot\vec n\;\chi\,
\Phi_0(\epsilon,r,\lambda)\ ,
\label{154}
 \eeq
%
 where the parameter $\lambda= 2\,a^2(\alpha)/\epsilon^2$ and the functions $\Phi_{0,1}$
are defined as
%
 \beq
\Phi_0(\epsilon,r,\lambda) =
\frac{1}{4\pi}\int\limits_{0}^{\infty}dt\,
\frac{\lambda}{{\rm sh}(\lambda t)}\,
{\rm exp}\left[-\ \frac{\lambda\epsilon^2 r^2}{4}\,
{\rm cth}(\lambda t) - t\right]\ ,
\label{160}
 \eeq
%
%
 \beq
\vec\Phi_1(\epsilon,r,\lambda) =
\frac{\epsilon^2\vec r}{8\pi}\int\limits_{0}^{\infty}dt\,
\left[\frac{\lambda}{{\rm sh}(\lambda t)}\right]^2\,
{\rm exp}\left[-\ \frac{\lambda\epsilon^2 r^2}{4}\,
{\rm cth}(\lambda t) - t\right]\ .
\label{162}
 \eeq
%
Here the functions $sh(x)$ and $cth(x)$ represent the hyperbolic sine and hyperbolic
cotangent, respectively.

The parameter $\lambda$ in Eqs.~(\ref{152}) and (\ref{154}) is responsible for the onset of the $q-\bar q$ interaction. In the limit of vanishing interaction $\lambda\to 0$ (i.e. $Q^2\to\infty$, $\alpha$ is fixed, $\alpha\not=0$ or $1$), Eqs.~(\ref{152}) and (\ref{154}) produce the well known perturbative expressions of Eq.~(\ref{122}). Then the matrix element (\ref{120}) contains the perturbative LC wave functions squared with the following form for T and L polarizations,
%
 \beq
\Bigl |\Psi^{T}_{q\bar q}(\vec r,\alpha,Q^2)\,\Bigr |^2 =
\frac{2\,N_C\,\alpha_{em}}{(2\pi)^2}\,
\sum_{f=1}^{N_f}\,Z_f^2
\left[m_f^2\,K_0(\epsilon,r)^2
+ [\alpha^2+(1-\alpha)^2]\,\epsilon^2\,K_1(\epsilon\,r)^2\right]\ ,
\label{197a}
 \eeq
%
and
%
 \beq
\Bigl |\Psi^{L}_{q\bar q}(\vec r,\alpha,Q^2)\,\Bigr |^2 =
\frac{8\,N_C\,\alpha_{em}}{(2\pi)^2}\,
\sum_{f=1}^{N_f}\,Z_f^2
\,Q^2\,\alpha^2(1-\alpha)^2\,
K_0(\epsilon\,r)^2\ ,
\label{197b}
 \eeq
%
where $K_1$ is the modified Bessel function,
$K_1(z) = -~d K_0(z)/dz$.

Using Eq.~(\ref{120}) for the total photoabsorption cross section including both polarizations T and L, one can obtain,
%
\beq
  \sigma_{tot}^{\gamma^*N}(x_{Bj},Q^2) =
  \sigma_T^{\gamma^*N}(x_{Bj},Q^2) +
  \tilde{\varepsilon}\,
  \sigma_L^{\gamma^*N}(x_{Bj},Q^2)\, ,
\label{198s1}
\eeq
%
with the photon polarization $\tilde{\varepsilon} = 1$ and
%
\beqn
 \sigma^{\gamma^*N}_T(x_{Bj},Q^2)
 &=&
 \int d^2 r \int_0^1 d\a \Bigl |\Psi_{q \bar q}^{T}(\vec{r},\a,Q^2)\Bigr |^2
 \,\sqqbar{(r)} \nonumber\\
 &&
 \hspace*{-3.5cm}
   =\frac{2N_C\a_{em}}{2\pi} \int_0^\infty r \,dr
   \int_0^1 d\a \sum_f Z_f^2 \left[(\a^2 + (1-\a)^2)\,\epsilon^2 K_1^2(\epsilon r) + m_f^2K_0^2(\epsilon r)\,\right]\sqqbar(r),
 \\
  \sigma^{\gamma^*N}_L(x_{Bj},Q^2)
  &=&
  \int d^2 r \int_0^1 d\a \Bigl |\Psi_{q \bar q}^{L}(\vec{r},\a,Q^2)\Bigr |^2
  \,\sqqbar{(\vec{r})}\nonumber \\
  &=&
 \frac{2 N_C\a_{em}}{2\pi} \int_0^\infty r \,dr \int_0^1
 d\a \sum_f Z_f^2\, 4\, Q^2\, \a^2(1-\a)^2\, K_0^2(\epsilon r)\,
 \sqqbar(r).
  \label{198s2}
\eeqn
%

Then the corresponding structure functions can be expressed as
%
\beqn
  F_1(\xBj,Q^2)
  &=&
  \frac{Q^2}{4\pi^2\a_{em}}
  \frac{\sigma_T^{\gamma^*N}(\xBj,Q^2)}{2\,\xBj},
  \label{198s3}\\
  F_2(\xBj,Q^2)
  &=&
  \frac{Q^2}{4\pi^2\a_{em}}
 \Bigl [\sigma_T^{\gamma^*N}(\xBj,Q^2) + \sigma_L^{\gamma^*N}(\xBj,Q^2)\Bigr ]\, , \label{198s4}
\eeqn
%
leading finally to the standard differential cross section for deep inelastic scattering,
%
\beqn
  \frac{d^2\sigma}{d\xBj dQ^2} = \frac{4\,\pi\,\alpha_{EM}^2}{Q^4}
  \left\{  \left(1 - y -
  \frac{\xBj^2\,y^2\,m_N^2}{Q^2}\right) \frac{F_2(\xBj,Q^2)}{\xBj}
  + y^2\,F_1(\xBj,Q^2)   \right\}\,.
\eeqn
%

Including the nonperturbative (npt) $q-\bar q$ interaction, one can obtain, instead of Eqs.~(\ref{197a}) and (\ref{197b}), the following expressions for the wave functions squared,
%
 \beq
\Bigl |\Psi^{T}_{npt}(\vec r,\alpha,Q^2)\,\Bigr |^2 =
{2\,N_C\,\alpha_{em}}\,
\sum_{f=1}^{N_f}\,Z_f^2
\left[m_f^2\,\Phi_0^2(\epsilon,r,\lambda)
+ [\alpha^2+(1-\alpha)^2]\,\bigl |\vec\Phi_1(\epsilon,r,\lambda)\,
\bigr |^2\,\right]\ ,
\label{199a}
 \eeq
%
and
%
 \beq
\Bigl |\Psi^{L}_{npt}(\vec r,\alpha,Q^2)\,\Bigr |^2 =
{8\,N_C\,\alpha_{em}}\,
\sum_{f=1}^{N_f}\,Z_f^2
\,Q^2\,\alpha^2(1-\alpha)^2\,
\Phi_0^2(\epsilon,r,\lambda)\ .
\label{199b}
 \eeq
%

As we have already mentioned above, within the LC formalism the energy resp. Bjorken-$\xBj$ dependence of the dipole cross section $\sigma_{q\bar q}(\vec{r},s)$ resp. $\sigma_{q\bar q}(\vec{r},\xBj)$ (see Eq.~(\ref{120})) accounts for the effect of higher Fock states $|q\bar qG\ra$, $|q\bar q2G\ra$, etc., which are contained in the photon wave function. Such an energy dependence of the dipole cross section is naturally included in various dipole models, like GBW \cite{GolecBiernat:1998js,Kowalski:2006hc}, KST \cite{Kopeliovich:1999am}, BGBK \cite{Bartels:2002cj} and IP-Sat \cite{Rezaeian:2012ji}, used in our calculations.


%
%
%
\subsection{Quark shadowing}
\label{quark-shad}
%
%
%

Now, we will switch on the nuclear targets and study the propagation of different Fock components of the photon in a nuclear matter. The derivation of the formula for nuclear shadowing, treating only the first shadowing correction for the lowest $|q\bar q\ra$ Fock state $\Delta\sigma_{tot}(\xBj,Q^2) = \Delta\sigma_{tot}^{q\bar q}(\xBj,Q^2)$ in Eq.~(\ref{110}), can be found
in \cite{Raufeisen:1998rg} and reads:
%
 \beq
\Delta\sigma_{tot}(\xBj,Q^2) = \frac{1}{2}~\Re~\int d^2 b
\int_{-\infty}^{
\infty} dz_1 ~\rho_{A}(b,z_1) \int_{z_1}^{\infty} dz_2~
\rho_A(b,z_2)
\int_{0}^{1} d\alpha ~A(z_1,z_2,\alpha)\, ,
\label{230}
 \eeq
%
where
%
 \beq
A(z_1,z_2,\alpha)
= \int d^2 r_2 \Psi^{*}_{q\bar q}(\vec{r_2},\alpha,Q^2)
\sigma_{q\bar q}(r_2,s) \int d^2 r_1
G_{q\bar q}(\vec{r_2},z_2;\vec{r_1},z_1)
\sigma_{q\bar q}(r_1,s)
\Psi_{q\bar q}(\vec{r_1},\alpha,Q^2)\, .
\label{240}
 \eeq
%
As soon as the nonpertubative interaction effects between the $q$ and $\bar q$ are explicitly included, one should use in Eq.~(\ref{240}) the LC wavefunctions $\Psi_{npt}(\vec r,\alpha,Q^2)$ instead of $\Psi_{q\bar q}(\vec r,\alpha,Q^2)$.

In Eq.~(\ref{230}) the variable $\rho_{A}(b,z)$ represents the nuclear density function defined at the point with longitudinal coordinate $z$ and impact parameter $\vec{b}$.
%
\begin{figure}[!htb]
   \includegraphics[angle=270,scale=0.5]{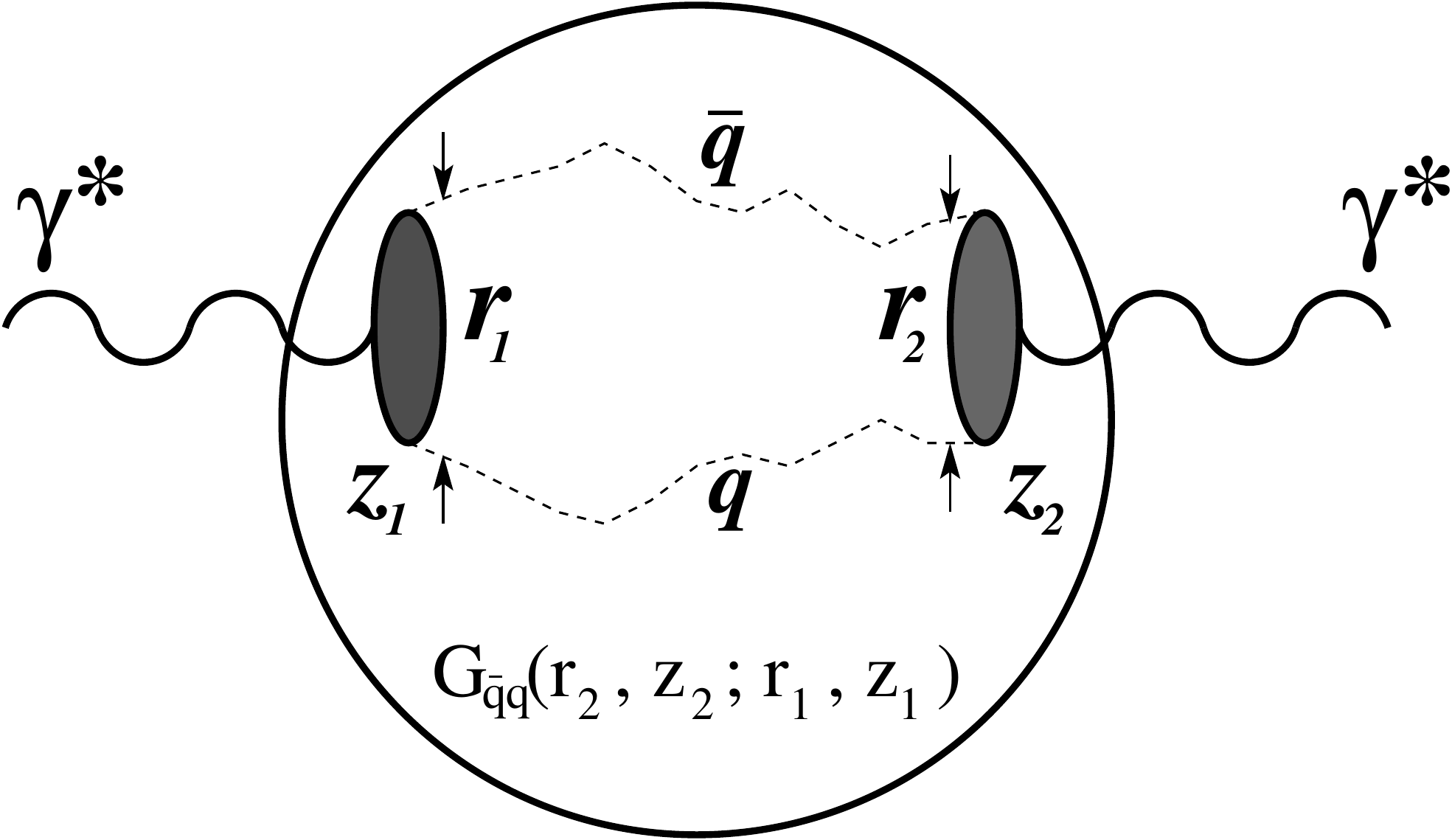}\centering
  \caption{
       A cartoon \cite{Kopeliovich:2001xj,Nemchik:2003wx,Kopeliovich:2008ek} for the first shadowing term $\Delta\sigma_{tot}(\xBj,Q^2) = \Delta\sigma_{tot}(q\bar q)$ in Eq.~(\ref{110}). The Green function $G_{q\bar q}(\vec{r_2},z_2;\vec{r_1},z_1)$ describes the propagation of the $q\bar q$ pair through the nucleus, which results from the summation over different paths of the $q\bar q$ pair.
           }
 \label{shad-qq}
\end{figure}
%

The first shadowing correction from the lowest $q\bar q$ component of the photon $\Delta\sigma_{tot}(\xBj,Q^2) = \Delta\sigma_{tot}^{q\bar q}(\xBj,Q^2)$ in (\ref{110}) is illustrated in Fig.~\ref{shad-qq} \cite{Kopeliovich:2001xj,Nemchik:2003wx,Kopeliovich:2008ek}. At the point with longitudinal coordinate $z_1$, the initial photon produces diffractively the $q\bar q$ pair ($\gamma^*N\to q\bar qN$) with initial transverse separation $\vec{r_1}$. The $q\bar q$ pair then propagates through the nucleus along arbitrary curved trajectories, which are summed over, and arrives at the point with longitudinal coordinate $z_2$ and with final transverse separation $\vec{r_2}$. The initial and final separations are controlled by the LC wave function of the $|q\bar q\ra$ Fock component of the photon $\Psi_{q\bar q}(\vec{r},\alpha,Q^2)$. During propagation through the nucleus, the $q\bar q$ pair interacts with bound nucleons via the dipole cross section $\sigma_{q\bar q}(r,s)$, which depends on the local transverse separation $\vec{r}$. The Green function $G_{q\bar q}(\vec{r_2},z_2;\vec{r_1},z_1)$ describes the propagation of the $q\bar q$ pair from longitudinal coordinate $z_1$ to $z_2$.

Similarly as in a vacuum,
the propagation of the $|q\bar q\ra$ Fock component of the photon in a nuclear medium is described
by the Green function $G_{q\bar q}(\vec{r_2},z_2;\vec{r_1},z_1)$ satisfying again the time-dependent two-dimensional Schr\"odinger equation, Eq.~(\ref{135}). However, here the potential $V_{q\bar q}(z_2,\vec r_2,\alpha)$ additionally acquires an imaginary part, which is responsible for the attenuation of the $q\bar q$ photon fluctuation in the medium and has the following form
%
 \beq
\Im V_{q\bar q}(z_2,\vec r,\alpha) = -
\frac{\sigma_{q\bar q}(\vec r,s)}{2}\,\rho_{A}({b},z_2)\, .
\label{250}
 \eeq
%

As was already mentioned above, only the harmonic oscillator potential $V_{q\bar q}(r)\propto r^2$ allows to solve Eq.~(\ref{135}) analytically. This corresponds to a quadratic approximation
also for $\Im V_{q\bar q}(z_2,\vec r_2,\alpha)$ and, consequently, one should take the dipole cross section of the form,
%
 \beq
\sigma_{q\bar q}(r,s) = C(s)\,r^2\ ,
\label{260}
 \eeq
%
and uniform nuclear density
%
 \beq
\rho_A(b,z) = \rho_0~\Theta(R_A^2-b^2-z^2)\, ,
\label{270}
 \eeq
%
with $R_A$ representing the nuclear radius. In this case the solution of Eq.~(\ref{135}) leads to the same form as that in Eq.~(\ref{142}), except that one should replace $\omega\Longrightarrow \Omega$ and $a^2(\alpha)\Longrightarrow
b(\alpha)$, where
%
\beq
\Omega =
\frac{b(\alpha)}{\nu \alpha (1 - \alpha)}
=
\frac{
\sqrt{a^4(\alpha) - i\,\rho_A(b,z)\,\nu\,\alpha\,(1 - \alpha)\,C(s)}}
{\nu\,\alpha\,(1 - \alpha)}\, .
\label{280}
\eeq
%

The value of the energy dependent factor $C(s)$ in Eqs.~(\ref{260}) and (\ref{280}) can be determined by the procedure described in Refs.~\cite{Kopeliovich:2000ra,Kopeliovich:2001xj,Nemchik:2003wx}. Here, the factor $C(s)$ was adjusted obtaining the same magnitude of nuclear shadowing in DIS employing the approximation Eq.~(\ref{260}) as well as the realistic parametrization of the dipole cross section (GBW \cite{GolecBiernat:1998js,Kowalski:2006hc}, KST \cite{Kopeliovich:1999am}, BGBK \cite{Bartels:2002cj} and IP-Sat \cite{Rezaeian:2012ji}) in the high energy limit, $l_c\gg R_A$, when the Green function acquires the simple form (see Eq.~(\ref{330})). This leads to the following relation,
%
\beqn
\frac
{\int d^2\,b\,\int d^2\,r\,\Bigl
|\Psi_{q\bar q}(\vec{r},\alpha,Q^2)\Bigr |^2\,
\left\{1 - \exp\,\Bigl[ - \frac{1}{2}\,C(s)\,r^2\,
T_A(b)\Bigr]\,\right\}}
{\int d^2\,r\,
\Bigl |\Psi_{q\bar q}(\vec{r},\alpha,Q^2)\Bigr |^2\,
C(s)\,r^2}
\nonumber \\
=
\frac{\int d^2\,b\,\int d^2\,r\,\Bigl
|\Psi_{q\bar q}(\vec{r},\alpha,Q^2)\Bigr |^2\,
\left\{1 - \exp\,\Bigl[ - \frac{1}{2}\,\sigma_{q\bar q}(r,s)\,
T_A(b)\Bigr]\,\right\}}
{\int d^2\,r\,
\Bigl |\Psi_{q\bar q}(\vec{r},\alpha,Q^2)\Bigr |^2\,
\sigma_{q\bar q}(r,s)}\, ,
\label{224}
\eeqn
%
where $T_A(b) = \int_{-\infty}^{\infty}\,dz\,\rho_A(b,z)$ is the nuclear thickness function calculated with the realistic Wood-Saxon form of the nuclear density, with parameters taken from \cite{DeJager:1987qc}. Analogically, the value $\rho_0$ of the uniform nuclear density Eq.~(\ref{270}) was fixed using the following relation \cite{Kopeliovich:2000ra,Kopeliovich:2001xj,Nemchik:2003wx}
%
\beq
\int\,d^2\,b\,\Biggl [1 - exp\,\Biggl ( - \sigma_0\,\rho_0
\,\sqrt{R_A^2 - b^2}\,\Biggr )\,\Biggr ] =
\int\,d^2\,b\,\Biggl [1 - exp\,\Biggl ( -
\frac{1}{2}\,\sigma_0\,T_A(b)\,\Biggr )\,\Biggr ]\, .
\label{228}
\eeq
%

Treating the shadowing correction for the lowest $|q\bar q\ra$ Fock component of the photon, the corresponding formula for the coherence length is given by Eq.~(\ref{10}), where the effective mass for the $q\bar q$ pair depends on the LC variable $\alpha$ and reads
%
\beq
M_{q\bar q}^2
\equiv
M_{q\bar q}^2(\alpha)
=
\frac{
m_q^2 + p_T^2}{\alpha(1-\alpha)}\, .
\label{effmas-qq}
\eeq
%
As was already discussed above, the CL effect is naturally included in the Green function formalism via the longitudinal momentum transfer of its minimal magnitude $q_L^{min} = 1/l_c^{max} = \varepsilon^2/(2\nu \alpha(1-\alpha))$ (see the second line of Eq.~(\ref{142})). Then, depending on the value of the CL, one can distinguish two regimes of the nuclear shadowing in DIS:

\textbf{(i)}
The first regime represents the general case with no restrictions for the CL $l_c$ and can be applied at any energy. At energies when $l_c\lsim R_A$, one has to take into account the variation of the transverse size $r$ during propagation of the $q\bar q$ pair through the nucleus. However, such a variation is naturally included in the Green function formalism presented above. Then the corresponding total photoabsorption cross section on a nucleus, after summation over all flavors, colors, helicities and spin states and expressed as a sum over T and L polarizations, $\sigma^{\gamma^*A} = \sigma_T^{\gamma^*A} + \epsilon'\,\sigma_L^{\gamma^*A}$, assuming the photon polarization $\epsilon'=1$, is given by the following expression \cite{Zakharov:1998sv,Raufeisen:1998rg,Nemchik:2003wx,Kopeliovich:2008ek},
%
\beqn
\sigma^{\gamma^*A}_{tot}(\xBj,Q^2) &=&
A\,\sigma^{\gamma^*N}_{tot}(\xBj,Q^2) - \Delta\,\sigma_{tot}^{q\bar q}(\xBj,Q^2)
\nonumber \\
&=& A\,\int\,d^2r\,\int_{0}^{1}\,d\alpha\,\sigma_{q\bar q}(r,s)
\,\Biggl (\Bigl |\Psi^T_{q\bar q}(\vec{r},\alpha,Q^2)\Bigr |^2 +
\Bigl |\Psi^L_{q\bar q}(\vec{r},\alpha,Q^2)\Bigr |^2\Biggr )
\nonumber \\
&-& \frac{N_C\,\alpha_{em}}{(2\pi)^2}\,\sum_{f=1}^{N_f}\,Z_f^2\,Re\,
\int\,d^2b\,\int_{-\infty}^{\infty}\,dz_1\,\int_{z_1}^{\infty}\,
dz_2\,\int_{0}^{1}\,d\alpha\,\int\,d^2r_1\,\int\,d^2r_2
\nonumber \\
&&\times\,\rho_A(b,z_1)\,\rho_A(b,z_2)\,\sigma_{q\bar q}(r_2,s)\,
\sigma_{q\bar q}(r_1,s)
\nonumber \\
&&\times\,\Biggl\{\Bigl[\,\alpha^2 + (1 - \alpha)^2\,\Bigr]
\,\epsilon^2
\,\frac{\vec{r_1}\,\cdot\,\vec{r_2}}{r_1\,r_2}\,
K_1(\epsilon\,r_1)\,K_1(\epsilon\,r_2)
\label{320}
\\
&&\,\,\,\,\,\,\,\,\,\,\,\,\,\,\,
 + \,\Bigl[\,m_f^2 + 4\,Q^2\,\alpha^2\,(1 - \alpha)^2\,\Bigr]\,
K_0(\epsilon\,r_1)\,K_0(\epsilon\,r_2)\Biggr\}\,
G_{q\bar q}(\vec{r_2},z_2;\vec{r_1},z_1) \, ,
\nonumber
\eeqn
%
where the photon LC wave functions squared $\Bigl |\,\Psi^{T,L}_{q\bar q}(\vec{r},\alpha,Q^2)\,\Bigr |^2$ are given by Eqs.~(\ref{197a}) and (\ref{197b}).

The Eq.~(\ref{320}) has the following modified form taking into account the nonperturbative interaction effects between $q$ and $\bar q$ of the virtual photon,
%
\beqn
\sigma^{\gamma^*A}_{npt}(\xBj,Q^2) &=&
A\,\sigma^{\gamma^*N}_{npt}(\xBj,Q^2) - \Delta\,\sigma_{npt}^{q\bar q}(\xBj,Q^2)
\nonumber \\
&=& A\,\int\,d^2r\,\int_{0}^{1}\,d\alpha\,\sigma_{q\bar q}(r,s)
\,\Biggl (\Bigl |\Psi^T_{npt}(\vec{r},\alpha,Q^2)\Bigr |^2 +
\Bigl |\Psi^L_{npt}(\vec{r},\alpha,Q^2)\Bigr |^2\Biggr )
\nonumber \\
&-& {N_C\,\alpha_{em}}\,\sum_{f=1}^{N_f}\,Z_f^2\,Re\,
\int\,d^2b\,\int_{-\infty}^{\infty}\,dz_1\,\int_{z_1}^{\infty}\,
dz_2\,\int_{0}^{1}\,d\alpha\,\int\,d^2r_1\,\int\,d^2r_2
\nonumber \\
&&\times\,\rho_A(b,z_1)\,\rho_A(b,z_2)\,\sigma_{q\bar q}(r_2,s)\,
\sigma_{q\bar q}(r_1,s)
\nonumber \\
&&
\times
\,\Biggl\{\Bigl[\,\alpha^2 + (1 - \alpha)^2\,\Bigr]
\,
\vec\Phi_1(\epsilon\,,r_1,\lambda)\cdot\vec\Phi_1(\epsilon\,,r_2,\lambda)
\label{325}
\\
&&\,\,\,\,\,\,\,
 + \,\Bigl[\,m_f^2 + 4\,Q^2\,\alpha^2\,(1 - \alpha)^2\,\Bigr]\,
\Phi_0(\epsilon\,,r_1,\lambda)\,\Phi_0(\epsilon\,,r_2,\lambda)\Biggr\}\,
G_{q\bar q}(\vec{r_2},z_2;\vec{r_1},z_1) \, ,
\nonumber
\eeqn
%
where $\Bigl |\,\Psi^{T,L}_{npt}(\vec{r},\alpha,Q^2)\,\Bigr |^2$ are
given by Eqs.~(\ref{199a}) and (\ref{199b}).

\textbf{(ii)}
In the high energy limit of long coherence length (LCL), $l_c\gg R_A$, the transverse separation $r$ between $q$ and $\bar q$ does not vary during propagation through the nucleus. The corresponding eikonal formula, frequently used in the literature, can be obtained as a limiting case of the Green function formalism, when the Green function takes a simple form,
%
\beq
G_{q\bar q}(b;\vec{r_2},z_2;\vec{r_1},z_1)|_{\nu\to\infty} =
\delta(\vec{r_2}-\vec{r_1})\,\exp\Biggl[ - \frac{1}{2}\,
\sigma_{q\bar q}(r_2,s)\,\int_{z_1}^{z_2}\,dz\,\rho_A(b,z)\Biggr]\, ,
\label{330}
\eeq
%
neglecting the kinetic term in Eq.~(\ref{135}) in the high energy limit $\nu\rightarrow \infty$.

Taking into account nonperturbative $q-\bar q$ interaction effects and substituting the Green function of the form (\ref{330}) into Eq.~(\ref{325}), one can obtain the following simple formula:
%
\beqn
\sigma^{\gamma^*A}_{npt}(\xBj,Q^2) &=&
2\,\int\,d^2b\,\int\,d^2r\,\int_0^1\,d\alpha
\left\{1 - \exp\,\Bigl[ - \frac{1}{2}\,\sigma_{\bar qq}(r,s)\,
T_A(b)\Bigr]\,\right\} \nonumber\\
&&\times\,{2\,N_C\,\alpha_{em}}\,\sum_{f=1}^{N_f}\,Z_f^2\,
\Biggl\{\Bigl[\,\alpha^2 + (1 - \alpha)^2\,\Bigr]
\,\Bigl |\vec\Phi_1(\epsilon\,,r,\lambda)\Bigr |^2\,
\label{335}
\\
&&\,\qquad\qquad\qquad\qquad
 + \,\Bigl[\,m_f^2 + 4\,Q^2\,\alpha^2\,(1 - \alpha)^2\,\Bigr]\,
\Phi_0^2(\epsilon\,,r,\lambda)\,\Biggr\}\, .
\nonumber
\eeqn
%
Here, to be more precise, one should replace the factor
$1 - \exp\,\bigl[ - \sigma_{\bar qq}(r,s)\,T_A(b)\bigr]/2$ by the expression,
$1 - \bigl[1 -\sigma_{\bar qq}(r,s)\,T_A(b)/(2 A) \bigr]^A$.

The explicit analytical form for the Green function Eq.~(\ref{142}) requires to use the quadratic harmonic oscillatory shape of the $q\bar q$ potential with the corresponding real part Eq.~(\ref{140}) as well as with the imaginary part given by Eq.~(\ref{250}) for the specific form of the dipole cross section Eq.~(\ref{260}) and the nuclear density function Eq.~(\ref{270}). However, in the general case, the evolution equation for the Green function has to be solved numerically and, consequently, one can use then the arbitrary phenomenological dipole models and realistic nuclear density functions. It was done for the first time in Ref.~\cite{Nemchik:2003wx} and is performed as well in the present paper.

The numerical solution of the Schr\"odinger equation, Eq.~(\ref{135}), for the Green function $G_{q\bar q}(\vec{r_2},z_2;\vec{r_1},z_1)$ with the initial condition, Eq.~(\ref{136}), is more effective performing the following substitutions \cite{Nemchik:2003wx},
%
\beqn
g_0(\vec{r_2},z_2;z_1,\lambda)
=
\int\,d^2r_1\,\Phi_0(\epsilon\,,r_1,\lambda)\,\sigma_{q\bar q}(r_1,s)\,
G_{q\bar q}(\vec{r_2},z_2;\vec{r_1},z_1),
\nonumber\\
\frac{\vec{r_2}}{r_2}\,g_1(\vec{r_2},z_2;z_1,\lambda)
=
\int\,d^2r_1\,\vec\Phi_1(\epsilon\,,r_1,\lambda)\,\sigma_{q\bar q}(r_1,s)\,
G_{q\bar q}(\vec{r_2},z_2;\vec{r_1},z_1)\, .
\label{350}
\eeqn
%
This leads to the following evolution equations for new functions $g_0(\vec{r_2},z_2;z_1,\lambda)$ and $g_1(\vec{r_2},z_2;z_1,\lambda)$
\cite{Nemchik:2003wx},
%
\beqn
i\frac{d}{dz_2}\,g_0(\vec{r_2},z_2;z_1,\lambda)\
\hspace*{12.0cm}
\nonumber\\
=
\left\{\frac{1}{2\,\nu\alpha(1-\alpha)}
\left[\epsilon^{2} -
\frac{\partial^2}{\partial\,r_2^2}
- \frac{1}{r_2}\,\frac{\partial}{\partial\,r_2}\right]
+ V_{q\bar q}(z_2,\vec r_2,\alpha)\right\}
g_0(\vec{r_2},z_2;z_1,\lambda),\
\hspace*{1.2cm}
\nonumber \\
i\frac{d}{dz_2}\,g_1(\vec{r_2},z_2;z_1,\lambda)
\hspace*{12.0cm}
\nonumber\\
=
\left\{\frac{1}{2\,\nu\alpha(1-\alpha)}
\left[\epsilon^{2} -
\frac{\partial^2}{\partial\,r_2^2}
- \frac{1}{r_2}\,\frac{\partial}{\partial\,r_2}
+ \frac{1}{r_2^2}\right]
+ V_{q\bar q}(z_2,\vec r_2,\alpha)\right\}
g_1(\vec{r_2},z_2;z_1,\lambda)\ ,
\nonumber\\
\label{370}
 \eeqn
%
where the real and imaginary part of the LC potential $V_{q\bar q}(z,\vec{r},\alpha)$
is given by Eqs.~(\ref{140}) and (\ref{250}), respectively.
The boundary conditions for the modified Green functions $g_0$ and $g_1$ in Eq.~(\ref{370}) read,
%
\beqn
g_0(\vec{r_2},z_2;z_1,\lambda)|_{z_2=z_1}=
\Phi_0(\epsilon\,,r_2,\lambda)\,\sigma_{\bar qq}(r_2,s),
\nonumber\\
g_1(\vec{r_2},z_2;z_1,\lambda)|_{z_2=z_1}=
\hat\Phi_1(\epsilon\,,r_2,\lambda)\,\sigma_{\bar qq}(r_2,s)\, ,
\label{390}
\eeqn
%
where the function $\hat\Phi_1$ is related to $\vec\Phi_1$ from Eq.~(\ref{162}) as $\vec\Phi_1(\epsilon\,,r,\lambda) = \frac{\vec{r}}{r}\cdot\hat\Phi_1(\epsilon\,,r,\lambda)$.

Consequently, the expression (\ref{325}), which includes explicitly the nonperturbative
interaction effects between $q$ and $\bar q$, will be modified as follows,
%
\beqn
\sigma^{\gamma^*A}_{npt}(\xBj,Q^2) &=&
A\,\sigma^{\gamma^*N}_{npt}(\xBj,Q^2) - \Delta\,\sigma_{npt}^{q\bar q}(\xBj,Q^2)
\nonumber \\
&=& A\,\int\,d^2r\,\int_{0}^{1}\,d\alpha\,\sigma_{q\bar q}(r,s)
\,\Biggl (\Bigl |\Psi^T_{npt}(\vec{r},\alpha,Q^2)\Bigr |^2 +
\Bigl |\Psi^L_{npt}(\vec{r},\alpha,Q^2)\Bigr |^2\Biggr )
\nonumber \\
&-& {3\,\alpha_{em}}\,\sum_{f=1}^{N_f}\,Z_f^2\,\,Re\,
\int\,d^2b\,\int_{-\infty}^{\infty}\,dz_1\,\int_{z_1}^{\infty}\,
dz_2\,\int_{0}^{1}\,d\alpha\,\int\,d^2r_2
\nonumber \\
&\times&
\,\rho_A(b,z_1)\,\rho_A(b,z_2)\,\sigma_{q\bar q}(r_2,s)\,
\Biggl\{\Bigl[\,\alpha^2 + (1 - \alpha)^2\,\Bigr]\,
\hat\Phi_1(\epsilon\,,r_2,\lambda)\,g_1(\vec{r_2},z_2;z_1,\lambda)
\label{420}
\nonumber
\\
&&\,\,\,\,\,\,\,
 + \,\Bigl[\,m_f^2 + 4\,Q^2\,\alpha^2\,(1 - \alpha)^2\,\Bigr]\,
\Phi_0(\epsilon\,,r_2,\lambda)\,g_0(\vec{r_2},z_2;z_1,\lambda)\Biggr\}\, ,
\eeqn
%
with details of the algorithm for the numerical solution of Eqs.~(\ref{370}) presented in Ref.~\cite{Nemchik:2003wx}.
%
\begin{figure}[bht]
  \centering
  \includegraphics[scale=1.]{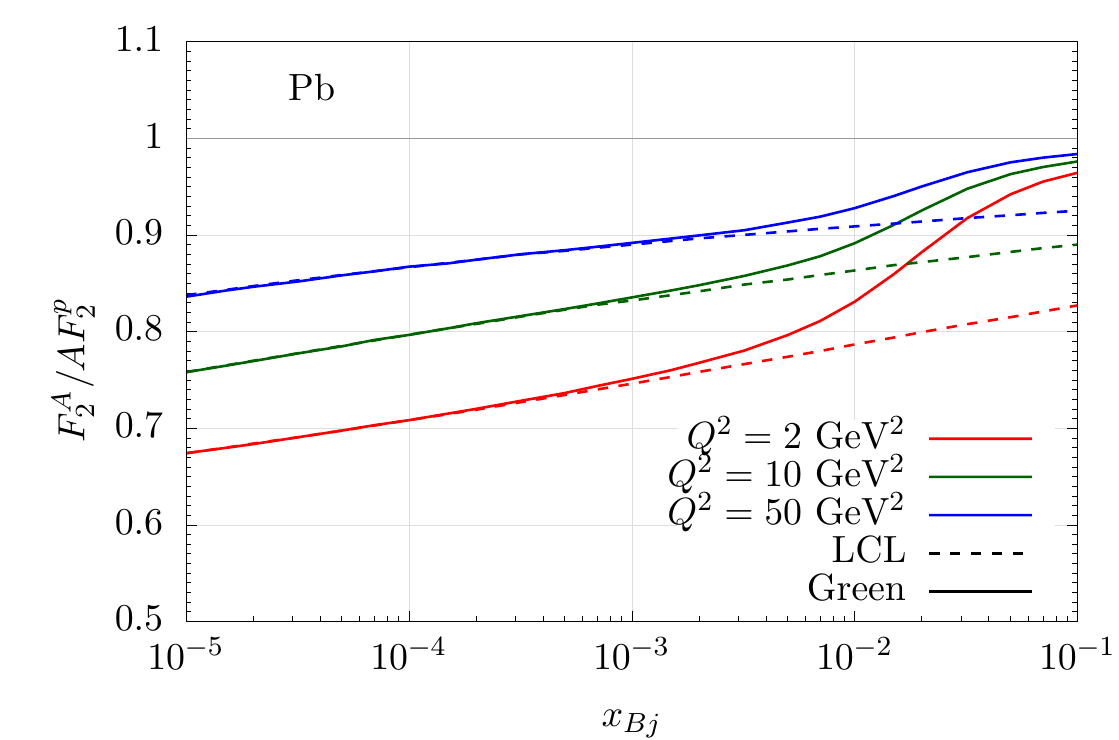}\centering
  \caption{
       Nuclear shadowing for the lowest $|q\bar q\ra$ Fock state as function of Bjorken $x_{Bj}$. Comparison of calculations in the LCL limit with more sophisticated Green function formalism at different fixed values of $Q^2$.
           }
  \label{fig:LCL}
\end{figure}
%

The most of recent studies of nuclear shadowing in DIS are based usually on calculations in the high energy (LCL) limit, $l_c\gg R_A$, also in kinematic regions when such a condition is not valid. The corresponding results overestimate thus the realistic shadowing at smaller energies when the coherence length Eq.~(\ref{10}) is comparable with the nuclear radius, $l_c\sim R_A$. Such a situation is demonstrated for the lowest $|q\bar q\ra$ Fock component in Fig.~\ref{fig:LCL} where we compare calculations of shadowing performed in the LCL limit, Eq.~(\ref{335}) (dashed lines) with the realistic case, Eq.~(\ref{420}) (solid lines), based on the Green function formalism. Such a comparison is presented as function of Bjorken $x_{Bj}$ (photon energy) for the lead target at different fixed values of the photon virtuality $Q^2$. One can see, that LCL calculations for the $|q\bar q\ra$ state can be
safely used for determination of the shadowing magnitude only at sufficiently small $x_{Bj}\lsim 10^{-3}$. Otherwise they overestimate the shadowing as is demonstrated as a difference between the solid and dashed lines at larger $x_{Bj}\gsim 10^{-3}$.

Note that inclusion of higher Fock states containing gluons makes a shift of such LCL limit towards smaller Bjorken $\xBj$ due to much larger values of the corresponding effective mass leading to much shorter coherence length (see Eq.~(\ref{10})).

Note that the lowest $|q\bar q\ra$ Fock component of the photon represents the highest twist shadowing correction \cite{Kopeliovich:2000ra}, and vanishes at large quark masses as $1/m_f^2$. This
does not happen for higher Fock states containing gluons, which will be discussed below.
Therefore, the contribution to nuclear shadowing from such Fock states represents the leading twist shadowing correction \cite{Kopeliovich:1999am,Kopeliovich:2002yv}. Moreover, a steep energy dependence of the dipole cross section $\sigma_{q\bar q}(r,s)$ (see Refs.~\cite{GolecBiernat:1998js,Kowalski:2006hc,Kopeliovich:1999am,Bartels:2002cj,Rezaeian:2012ji} for different dipole models, for example) especially at smaller dipole sizes $r$ causes a strong energy rise of both corrections.

%
%
%
\subsection{Gluon shadowing}
\label{glue-shad}
%
%
%

Within the LC dipole approach based on the Green function formalism, the first shadowing correction in Eq.~(\ref{112}) containing just one gluon corresponds to the Fock component  $|q\bar qG\ra$. In comparison with the $|q\bar q\ra$ state, such a fluctuation has a larger effective mass $M^2_{q\bar qG}$ (see Eq.~(\ref{effmass-qqG})) leading so to  a smaller coherence time given by Eq.~(\ref{10}). Consequently, the larger photon energy (smaller Bjorken $x_{Bj}$) is required for a manifestation of an analogical onset of shadowing effects as for the $|q\bar q\ra$ case. The detailed discussion about gluon shadowing correction as well as different interpretation of this effect in various reference frames can be found in Refs.~\cite{Kopeliovich:1999am,Kopeliovich:2001hf,Kopeliovich:2008ek}.

In the present paper, the gluon shadowing correction related to the $|q\bar qG\ra$ intermediate Fock component has been calculated analogically as presented in Refs.~\cite{Kopeliovich:1999am,Kopeliovich:2001hf,Kopeliovich:2008ek}. Here, the suppression factor $R_G$ is given as the ratio of gluon densities in nuclei and nucleon, %
\begin{equation}
\label{eq:dipole:gs:RGdef}
    R_G(x_{Bj},Q^2) = \frac{G_A(x_{Bj},Q^2)}{A\,G_N(x_{Bj},Q^2)}
    \sim 1 - \frac{\Delta \sigma_{tot}(q\bar qG)}{A\,\sigma_{tot}^{\gamma^* N} (x_{Bj},Q^2)},
\end{equation}
%
where the inelastic correction $\Delta\sigma_{tot}(q\bar qG)$ to the total cross section $\sigma_{tot}^{\gamma^*A}(x_{Bj},Q^2)$ (see Eq.~(\ref{110})) reads,
%
 \beqn
&&
\hspace*{-0.9cm}
\Delta\sigma_{tot}(q\bar qG)
=
{\rm Re}
\int\limits_{0}^{\infty} d^{\,2} b
\int\limits_{-\infty}^{\infty}
dz_2 \int\limits_{-\infty}^{z_2} dz_1\,
\rho_A(b,z_1)\,\rho_A(b,z_2)
\int d^{\,2}x_2\,d^{\,2}y_2\,d^{\,2}x_1\,d^{\,2}y_1 \int
d\alpha_q\,\frac{d\,\alpha_G}{\alpha_G}
\nonumber\\ &\times&
F^{\dagger}_{\gamma^*\to q\bar qG}
(\vec x_2,\vec y_2,\alpha_q,\alpha_G)\
G_{q\bar qG}(\vec x_2,\vec y_2,z_2;\vec x_1,\vec y_1,z_1)\
F_{\gamma^*\to q\bar qG}
(\vec x_1,\vec y_1,\alpha_q,\alpha_G)\, ,
\nonumber\\
\label{delta-sigma}
 \eeqn
%
where variables $\vec x$ and $\vec y$ represent the transverse distances from the gluon
to the quark and antiquark, respectively; $\alpha_q$ is the fraction
of the LC momentum of the $q\bar q$ carried by the quark;
$\alpha_G$ is the fraction of the photon momentum carried by the
gluon; and $G_{q\bar qG}(\vec x_2,\vec y_2,z_2;\vec
x_1,\vec y_1,z_1)$ is the LC Green function describing the
propagation of the $q\bar qG$ system between states with initial
longitudinal and transverse coordinates $z_1$ and $\vec x_1,\vec
y_1$, respectively, and the final coordinates $(z_2,\vec x_2,\vec
y_2)$.

The above Eq.~(\ref{delta-sigma}) contains also functions
$F_{\gamma^*\to q\bar qG}$, representing the amplitudes of diffractive
$q\bar qG$ production in a $\gamma^*N$ interaction \cite{Kopeliovich:1999am},
of the following form,
%
 \beqn
F_{\gamma^*\to q\bar qG}(\vec x,\vec y,\alpha_q,\alpha_G)
&=&
\frac{9}{8}\,
\Psi_{q\bar q}(\vec x -\vec y,\alpha_q)\,
\left[\Psi_{qG}\left(\vec x,\frac{\alpha_G}{\alpha_q}
\right) - \Psi_{\bar qG}
\left(\vec y,\frac{\alpha_G}{1-\alpha_q}\right)\,
\right]\nonumber\\
&\times&
\biggl[\sigma_{q\bar q}(x)+
\sigma_{q\bar q}(y) -
\sigma_{q\bar q}(\vec x - \vec y)\biggr]\ ,
\label{amplitude}
\eeqn
%
where $\Psi_{q\bar q}$ (see Eqs.~(\ref{152}) and (\ref{154}))
and $\Psi_{\bar qG}$ are the LC distribution
functions of the $q\bar q$ fluctuations of a photon and $qG$
fluctuations of a quark, respectively.
The latter includes nonperturbative interaction effects and has the following form \cite{Kopeliovich:1999am},
%
\beq
    \Psi_{qG}(r,\alpha) = \frac{2}{\pi}\,
    \frac{\sqrt{\alpha_S({Q}^2)}}
    {\sqrt{3}}\,
    \frac{\vec e_q\,\cdot \vec r}{r^2}
    \exp \left( - \frac{1}{2}\, \tilde b^2 (\alpha)\, r^2 \right)\, ,
\label{qG}
\eeq
%
where $\vec e_q$ is the quark polarization vector; and $\tilde b(\a) = b_0^2 + \a Q^2$, with $b_0^2 = (0.65)^2\,\GeV^2$. Here the nonperturbative parameter $b_0$ is related to the mean quark-gluon separation $r_0$ as $b_0\sim 1/r_0$.
%
\begin{figure}[!htb]
   \includegraphics[angle=0,scale=0.5]{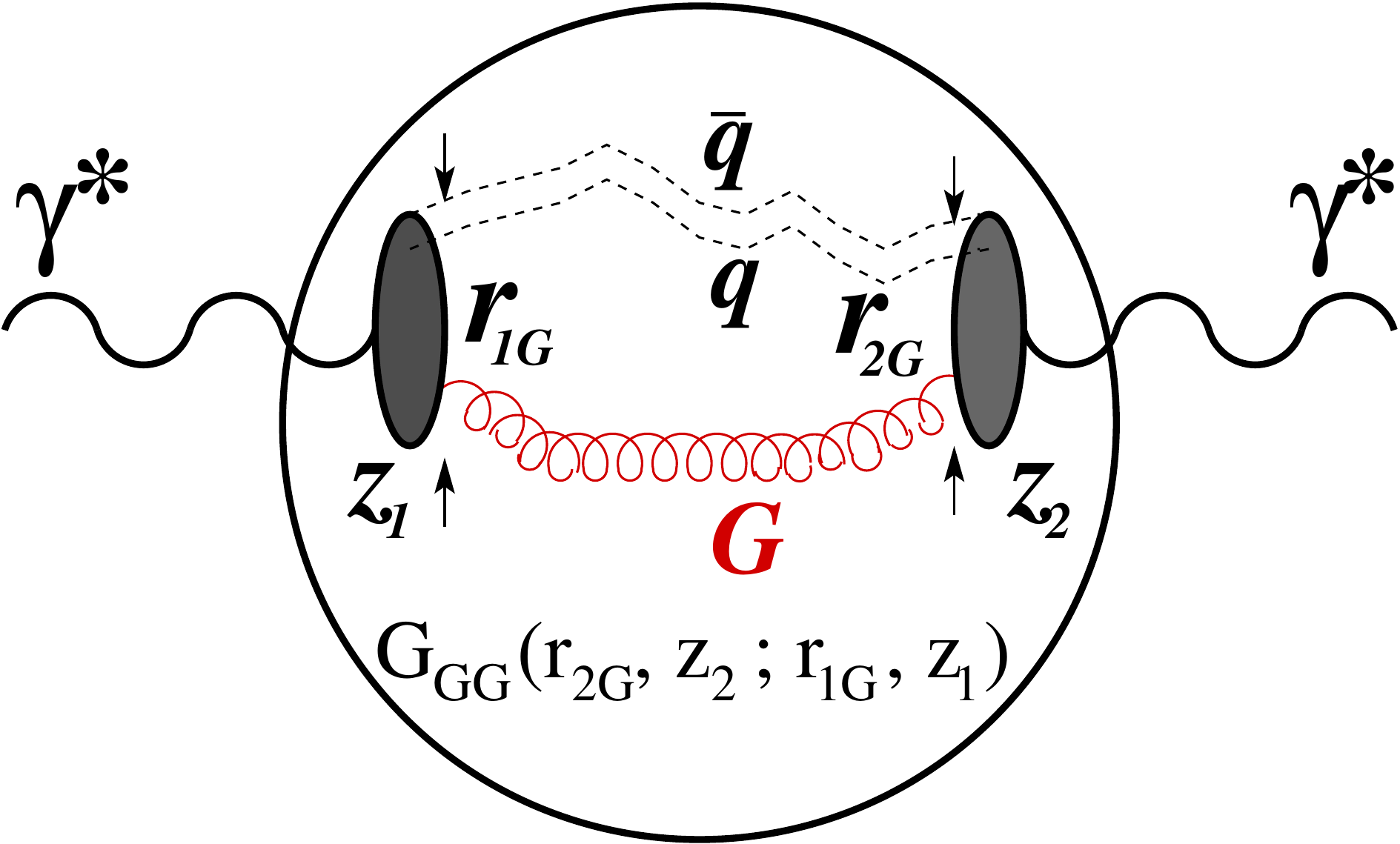}\centering
  \caption{
       A cartoon \cite{Kopeliovich:2016jjx} for the shadowing term $\Delta\sigma_{tot}(\xBj,Q^2) = \Delta\sigma_{tot}(q\bar qG)$. The Green function
       $G_{GG}(\vec{r}_{2G},z_2;\vec{r}_{1G},z_1)$ describes the propagation
       of the $q\bar qG$ system through the nucleus as a propagation of the
       effective gluon-gluon (color octet-octet) dipole neglecting the small transverse
       size of the color-octet $G\equiv  q\bar q$ fluctuation.
           }
 \label{shad-qqg}
\end{figure}
%

The LC Green function in Eq.~(\ref{delta-sigma}) describing the propagation of the three-body system $q\bar qG$ can be simplified without any significant impact on the accuracy
of gluon shadowing calculations.
As was discussed in Ref.~\cite{Kopeliovich:1999am} and is presented below in Sect.~\ref{npt} (see also Fig.~\ref{fig:Pb-npt}), at large $Q^2$ the calculation of shadowing corresponding to $|q\bar q\ra$ Fock state is based mainly on perturbative QCD. However, the nonperturbative effects can not be neglected for the higher $|q\bar qG\ra$ component. Consequently, for $Q^2\gg 1/r_0^2 = b_0^2\approx 0.45\,\GeV^2$ the mean $q\bar q$ transverse size squared (see Eq.~(\ref{130})) $\la r^2\ra_{q\bar q}\ll r_0^2$. In this case suppressing the intrinsic $q\bar q$ separation, i.e. assuming $\vec x = \vec y$, one can obtain a more simple form for the Green function, which describes now effectively the propagation of a two-body gluon-gluon (color octet-octet) dipole through a medium as is illustrated in Fig.~\ref{shad-qqg}.

All details of calculation can be found in Refs.~\cite{Kopeliovich:1999am,Kopeliovich:2001hf,Kopeliovich:2008ek}. In the present paper, we test for the first time how the magnitude of gluon shadowing correction is correlated with the shape of several specific models for the dipole cross section (see
Refs.~\cite{GolecBiernat:1998js,Kowalski:2006hc} for the GBW,
\cite{Kopeliovich:1999am} for the KST, \cite{Bartels:2002cj} for the BGBK, and \cite{Rezaeian:2012ji} for the IP-sat parametrization) used in our analyses.

The final expression for the gluon suppression factor (\ref{eq:dipole:gs:RGdef})
contains the factor C \cite{Kopeliovich:1999am,Kopeliovich:2001hf,Kopeliovich:2008ek}
inherent in the quadratic form of the dipole cross section (\ref{260}) which can be usually obtained as the first term of the Taylor series,
%
\begin{equation}\label{C0}
  C_0(x) =
  \left.\frac{d \sigma_{GG}(r,x)}{dr^2}\right|_{r^2=0} =
  \frac{9}{4}\,
  \left.\frac{d \sigma_{q \bar q}(r,x)}{dr^2}\right|_{r^2=0}\, .
\end{equation}
%

However, more realistic determination of the parameter $C\to C_{eff}$ follows from the asymptotic condition \cite{Kopeliovich:2001hf} similar to that for the $q\bar q$ component of the photon (\ref{224}),
%
\beqn
 \frac{\int d^2b\,\int d^2r \,
 \Bigl |\Psi_{qG}(r,\alpha_G)\Bigr|^2\,
 \Bigl\{1 - \exp\bigl [ -\frac{1}{2} C_{eff}(x,Q^2)\,T_A(b)\,r^2\bigr ]\Bigr\}}
 {
 \int d^2r \,
 \Bigl |\Psi_{qG}(r,\alpha_G)\Bigr|^2\,
 C_{eff}(x,Q^2)\,r^2} \nonumber\\
  =
 \frac{\int d^2b\,\int d^2r \,
 \Bigl |\Psi_{qG}(r,\alpha_G)\Bigr|^2\,
 \Bigl\{1 - \exp\bigl [ -\frac{9}{8}\,\sqqbar(r,x)\,T_A(b)\bigr ]\Bigr\}}
{
 \int d^2r \,
 \Bigl |\Psi_{qG}(r,\alpha_G)\Bigr|^2\,
 \frac{9}{4}\,\sqqbar(r,x)} \, ,
    \label{eq:eA:Ceff}
\eeqn
%
where
the dipole cross section $\sigma_{q\bar q}$ is sampled at the energy corresponding
to $x=x_{Bj}/\alpha_G$ with the prescription $x=min(x_{Bj}/\alpha_G,0.1)$ \cite{Kopeliovich:1999am,Kopeliovich:2001hf,Kopeliovich:2008ek} and
the LC quark-gluon wave function squared $|\Psi_{qG}|^2$ follows from Eq.~(\ref{qG}) and
has the following form,
%
\beq
    \Bigl |\Psi_{qG}(r,\alpha)\Bigr |^2 = \frac{4\alpha_S(Q^2)}{3\pi^2} \,
    \frac{\exp \left( -\tilde{b}^2(\alpha)\, r^2 \right)}{r^2}.
\eeq
%

%
\begin{figure}[htb]
  \centering
  \includegraphics[scale=1.4]{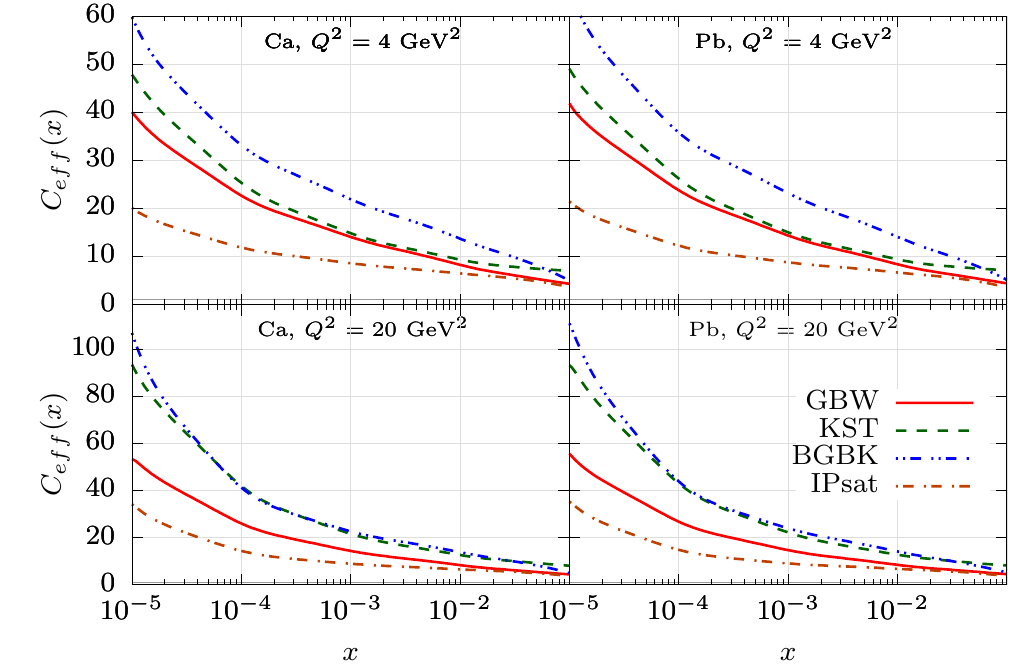}\centering
  \caption{
       The $x$-dependence of the factor $C_{eff}$ at fixed value $\xBj = 10^{-6}$ for the $Ca$ (left panels) and $Pb$ (right panels) target, respectively. The top and bottom panels correspond $Q^2 = 4\,\GeV^2$ and at $Q^2 = 20\,\GeV^2$, respectively. Determination of the factor $C_{eff}$ is performed for several phenomenological dipole cross sections as mentioned in the text.
          }
  \label{fig:Ceff}
\end{figure}
%

Results of extraction of dipole parameters $C_{eff}$ from the asymptotic
condition (\ref{eq:eA:Ceff}) are presented in Fig.~\ref{fig:Ceff} for the $Ca$ and $Pb$ targets.
Here we present also the sensitivity of such extraction to different dipole parametrizations
used in our analyses of nuclear shadowing, such as GBW
\cite{GolecBiernat:1998js,Kowalski:2006hc}, KST \cite{Kopeliovich:1999am},
BGBK \cite{Bartels:2002cj} and IP-sat \cite{Rezaeian:2012ji}.
Because at small transverse separations $r$ the dipole cross section is related to
the gluon structure function $G_N(x,Q^2)$ of the target nucleon as,
%
\beq
\sigma_{q\bar q}(r,x) = \frac{\pi^2}{3}\,\alpha_S(r)\,G_N(x,Q^2)\,r^2\, ,
\eeq
the growth of the factor $C_{eff}$ with energy is caused by the rise
of $G_N(x,Q^2)$ towards small $x$. Consequently, all panels of Fig.~\ref{fig:Ceff} also clearly demonstrate that uncertainties in determination of $C_{eff}$, caused by different dipole model parametrizations, rise towards smaller values of $x$.

%
\begin{figure}[htb]
  \centering
  \includegraphics[scale=1.4]{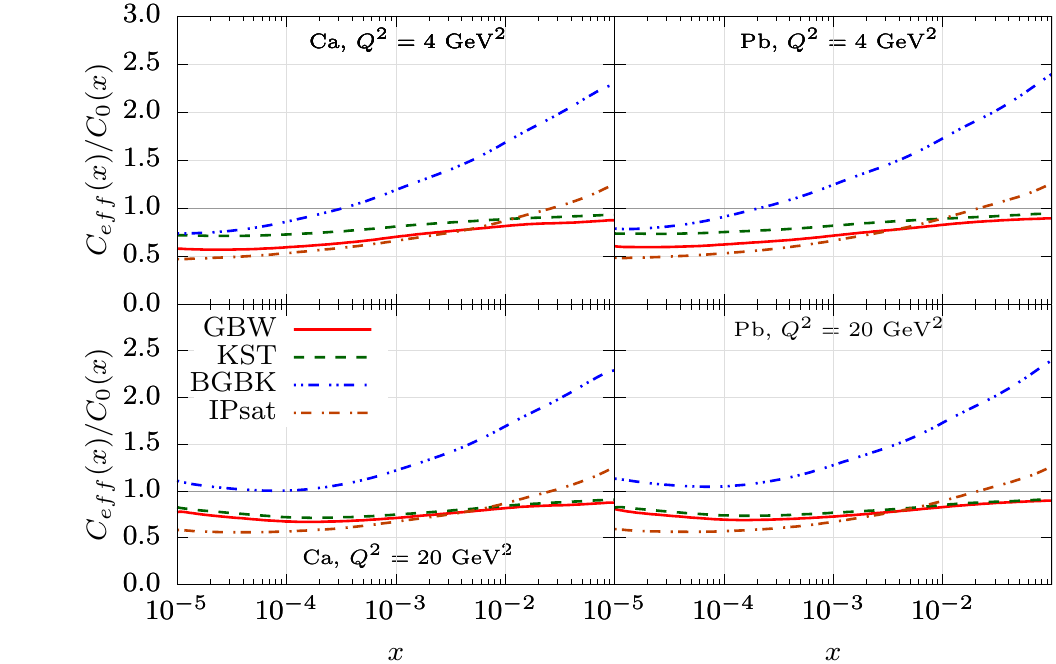}\centering
  \caption{
       The same as Fig.~\ref{fig:Ceff} but for
       the ratio of factors $C_{eff}/C_0$.
          }
  \label{fig:RatioCeffC0}
\end{figure}
%

The differences between standard determination of the $C$ parameter based on Eq.~(\ref{C0}) ($C\Rightarrow C_0$)
and more realistic extraction based on the asymptotic condition (\ref{eq:eA:Ceff}) ($C\Rightarrow C_{eff}$) are depicted
in Fig.~\ref{fig:RatioCeffC0} in terms of the $x$-dependent ratio
$C_{eff}/C_0$ for the $Ca$ and $Pb$ targets. Such a comparison has been performed again for different models of the dipole cross section denoted as GBW, KST, BGBK and IP-sat with corresponding references presented above. One can see that at small $x$, the factor $C_{eff} < C_0$, what leads to a smaller magnitude of gluon shadowing correction using more realistic determination of the factor $C_{eff}$.
Besides, the GBW, KST and IP-Sat dipole models give quite similar results, which differ significantly from the results based on BGBK model, especially at larger values of $x\gsim 10^{-4}$ and larger values of $Q^2$.

Within the LC dipole formalism, one can calculate the gluon shadowing correction
corresponding to the lowest Fock component $|q\bar qG\ra$. Since the inclusion of higher multigluon fluctuations is very complicated, their contribution to gluon shadowing can be effectively included eikonalizing the calculated factor $R_G(x_{Bj},Q^2)$ \cite{Kopeliovich:2001ee}. This leads to the following renormalization of the dipole cross section anywhere in expressions for the photoabsorption cross section,
%
\beq
\label{eq:dipole:gs:replace}
    \sqqbar(r,x) \Rightarrow \sqqbar(r,x)\,R_G(x,Q^2)\, .
\eeq
%

Consequently, the final formula (\ref{420}) for the nuclear total photoproduction cross section $\sigma_{npt}^{\gamma^* A}$, including nonperturbative $q-\bar q$ interaction effects and corrected also for the gluon shadowing effects, now reads,
%
\beqn
&&\sigma^{\gamma^*A}_{npt}(\xBj,Q^2,R_G)
=
\int d^2b\,T_A(b)\,R_G(b,\xBj,Q^2)\,\sigma^{\gamma^*N}_{npt}(\xBj,Q^2) - \Delta\,\sigma_{npt}^{q\bar q}(\xBj,Q^2,R_G)
\nonumber \\
&=&
\int d^2b\,T_A(b)\,
\int\,d^2r\,\int_{0}^{1}\,d\alpha\,\sigma_{q\bar q}(r,s)\,
R_G(b,\xBj,Q^2)\,
\Biggl (\Bigl |\Psi^T_{npt}(\vec{r},\alpha,Q^2)\Bigr |^2 +
\Bigl |\Psi^L_{npt}(\vec{r},\alpha,Q^2)\Bigr |^2\Biggr )
\nonumber \\
&-&
{3\,\alpha_{em}}\,\sum_{f=1}^{N_f}\,Z_f^2\,\,Re\,
\int\,d^2b\,\int_{-\infty}^{\infty}\,dz_1\,\int_{z_1}^{\infty}\,
dz_2\,\int_{0}^{1}\,d\alpha\,\int\,d^2r_2\,
\rho_A(b,z_1)\,\rho_A(b,z_2)\,
\nonumber \\
&\times&
\sigma_{q\bar q}(r_2,s)\,R_G(b,\xBj,Q^2)\,
\Biggl\{\Bigl[\,\alpha^2 + (1 - \alpha)^2\,\Bigr]\,
\hat\Phi_1(\epsilon\,,r_2,\lambda)\,h_1(\vec{r_2},z_2;z_1,\lambda,b)
\label{420rg}
\nonumber
\\
&&\hspace*{3.0cm}
 + \,\Bigl[\,m_f^2 + 4\,Q^2\,\alpha^2\,(1 - \alpha)^2\,\Bigr]\,
\Phi_0(\epsilon\,,r_2,\lambda)\,h_0(\vec{r_2},z_2;z_1,\lambda,b)\Biggr\}\, ,
\eeqn
%
where the modified Green functions
$h_0$ and $h_1$ are now defined as (compare with Eq.~(\ref{350})),
%
\beqn
h_0(\vec{r_2},z_2;z_1,\lambda,b)
=
\int\,d^2r_1\,\Phi_0(\epsilon\,,r_1,\lambda)\,\sigma_{q\bar q}(r_1,s)\,
R_G(b,\xBj,Q^2)\,G_{q\bar q}(\vec{r_2},z_2;\vec{r_1},z_1),
\nonumber\\
\frac{\vec{r_2}}{r_2}\,h_1(\vec{r_2},z_2;z_1,\lambda,b)
=
\int\,d^2r_1\,\vec\Phi_1(\epsilon\,,r_1,\lambda)\,\sigma_{q\bar q}(r_1,s)\,
R_G(b,\xBj,Q^2)\,G_{q\bar q}(\vec{r_2},z_2;\vec{r_1},z_1)\, ,
\nonumber\\
\label{350rg}
\eeqn
%
and satisfy the same Schr\"odinger equations as those given by Eq.~(\ref{370})
but with the following boundary conditions,
%
\beqn
h_0(\vec{r_2},z_2;z_1,\lambda,b)|_{z_2=z_1}=
\Phi_0(\epsilon\,,r_2,\lambda)\,\sigma_{\bar qq}(r_2,s)\,R_G(b,\xBj,Q^2),
\nonumber\\
h_1(\vec{r_2},z_2;z_1,\lambda,b)|_{z_2=z_1}=
\hat\Phi_1(\epsilon\,,r_2,\lambda)\,\sigma_{\bar qq}(r_2,s)\,R_G(b,\xBj,Q^2)\, .
\label{390rg}
\eeqn
%

Since we use the quadratic approximation of the dipole cross section (see Eq.~(\ref{C0})) in calculations of gluon shadowing correction, for large dipole separations we implement a restriction that the corresponding magnitude of the dipole cross section does not exceed maximal values of $\sigma_0$ inherent in the realistic saturated forms of the dipole cross sections analyzed in the present paper. This leads to the following condition,
%
\begin{equation}
\label{}
C_0\,  \bar{r}^2(b,\Delta z = z_2-z_1) \le \sigma_0,
\end{equation}
%
resulting into the following substitution,
%
\beq
\bar{r}^2(b,\Delta z)\Rightarrow
\frac{\sigma_0}{C_0}\,
\left\{1 - \exp\left [ - \frac{C_0\,\bar{r}^2(b,\Delta z)}{\sigma_0}\right]\right\}\, ,
\eeq
%
where the mean transverse size squared $\bar{r}^2(b,\Delta z)\ra$ reads,
%
\beq
\label{}
\bar{r}^2(b,\Delta z) =
\frac{
Re\,
\int_{0}^{1}\,d\alpha\,\int\,d^2r_2\,r^2(b,\Delta z)\,
\rho_A(b,z_1)\,\rho_A(b,z_2)\,\sigma_{q\bar q}(r_2,s)\,
R_G(b)\,
F(\alpha,R_2,z_1,z_2)
}
{
Re\,
\int_{0}^{1}\,d\alpha\,\int\,d^2r_2\,
\rho_A(b,z_1)\,\rho_A(b,z_2)\,\sigma_{q\bar q}(r_2,s)\,
R_G(b)\,
F(\alpha,r_2,z_1,z_2)
}
\eeq
%
with the function $F(\alpha,r_2,z_1,z_2)$ defined from Eq.~(\ref{420rg}),
%
\beqn
F(\alpha,r_2,z_1,z_2)
=
\Biggl\{\Bigl[\,\alpha^2 + (1 - \alpha)^2\,\Bigr]\,
\hat\Phi_1(\epsilon\,,r_2,\lambda)\,h_1(\vec{r_2},z_2;z_1,\lambda,b)\,
\nonumber
\\
 + \,\Bigl[\,m_f^2 + 4\,Q^2\,\alpha^2\,(1 - \alpha)^2\,\Bigr]\,
\Phi_0(\epsilon\,,r_2,\lambda)\,h_0(\vec{r_2},z_2;z_1,\lambda,b)\Biggr\}\, .
\eeqn
%

%
%
%
\section{Gluon shadowing from higher Fock components}
\label{glue-all}
%
%
%

As was mentioned in the previous Sect.~\ref{glue-shad}, the effect
of higher Fock states of the photon containing gluons has been included by means of
Eq.~(\ref{eq:dipole:gs:replace}) simulating thus effectively the interference effects in a nucleus.
Although the total virtual photoabsorption cross section
$\sigma_{tot}^{\gamma^*A}(x_{Bj},Q^2)$ can be expressed by Eq.~(\ref{110})
in terms of shadowing corrections decomposed over different Fock components
as is given by Eq.~(\ref{112}), the corresponding gluon shadowing factor $R_G$ in Eq.~(\ref{eq:dipole:gs:replace}) has been calculated taking only the $q\bar qG$ fluctuation.
In comparison to the $|q\bar q\ra$ state, such a one-gluon fluctuation has a significantly smaller coherence time (\ref{10}) due to its larger effective mass,
which reads \cite{Kopeliovich:2000ra},
%
\beq
M_{q\bar qG}^2\equiv
M_{q\bar qG}^2(\alpha,\alpha_G)
=
\frac{p_T^2}{\alpha_G (1-\alpha_G)} +
\frac
{M_{q\bar q}^2(\alpha)}{1-\alpha_G}
\gg
M_{q\bar q}^2
\, ,
\label{effmass-qqG}
\eeq
%
where $\alpha_G$ and $\alpha$ have been introduced above and $M_{q\bar q}^2$ is the effective mass of the $q\bar q$ fluctuation given by Eq.~(\ref{effmas-qq}).

The next two-gluon Fock component $|q\bar q 2G\ra$ has even much higher effective mass, $M_{q\bar q2G}^2\gg M_{q\bar qG}^2$ leading thus to a much shorter coherence time and, consequently, the shadowing correction terms $\Delta\sigma_{tot}^{q\bar q2G}$, $\Delta\sigma_{tot}^{q\bar q3G}$, ... , $\Delta\sigma_{tot}^{q\bar qnG}$ in Eq.~(\ref{112}) are negligibly small within the kinematic range accessible by the future experiments at EICs. 
For this reason, in the present paper we perform predictions for nuclear shadowing keeping only $|q\bar q\ra$ and $|q\bar qG\ra$ Fock states.

Finally, we would like to emphasize that the summation of all Fock components is inherent in Balitsky-Kovchegov (BK) equation \cite{Balitsky:1995ub,Kovchegov:1999yj}. Here, calculating shadowing in DIS off nuclear targets, all Fock states are treated in the high energy limit when the corresponding coherence length exceeds significantly the nuclear radius, i.e. $l_c^{q\bar q}, l_c^{q\bar qG}, ... ,l_c^{q\bar qnG}\gg R_A$. Consequently, BK equation can not lead to reliable predictions for nuclear shadowing especially in kinematic regions, studied in the present paper and covered by the future EICs,
where the coherence length for some Fock states, which are dominant for shadowing, can be comparable with the nuclear radius.

%
%
%
\section{Predictions for nuclear shadowing}
\label{results}
%
%
%

%
%
%
\subsection{Kinematic regions of planned electron-ion colliders}
\label{Sec:kin}
%
%
%

\begin{figure}[!htb]
  \centering
  \includegraphics[scale=1.]{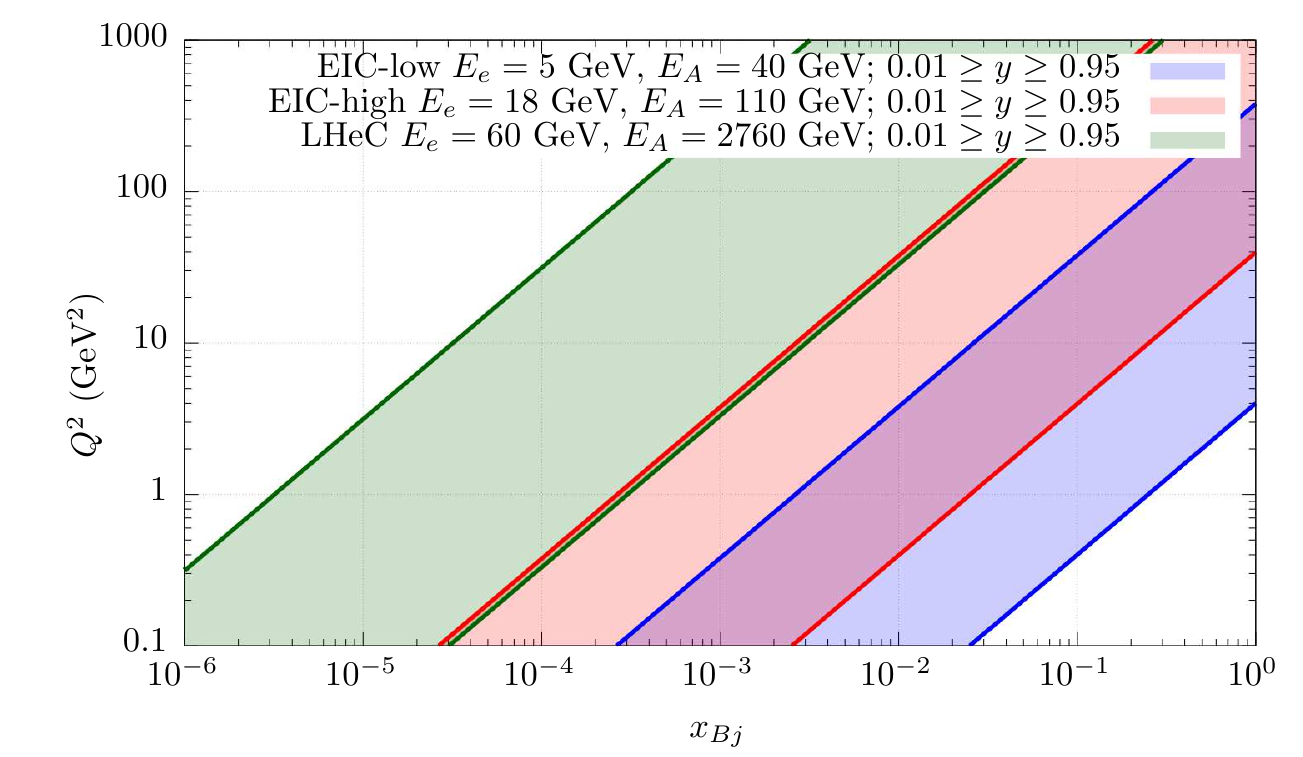}\centering
  \caption{
       The $Q^2$ vs $x_{Bj}$ ranges corresponding to kinematic regions covered by the future nuclear DIS experiments at several electron-ion colliders planned at RHIC in BNL (EIC) \cite{Accardi:2012qut,Aschenauer:2014cki,Aschenauer:2017jsk}, as well as at LHC (LHeC) in CERN \cite{AbelleiraFernandez:2012cc}.
       }
  \label{fig:Q2x}
\end{figure}

The kinematic regions covered by the future experiments at several electron-ion
colliders are presented in Fig.~\ref{fig:Q2x}.
%
%
Table~\ref{tab:EIC:energy} shows the energy ranges accessible by the future EIC experiment at RHIC in BNL, as well as at LHeC in CERN. In the case of EIC, the nuclear effects are expected to be investigated using mainly the gold target. However, other nuclear targets can be studied as different nuclear beams were set at RHIC in the past. At the LHeC, the measured nuclear target (lead) should remain the same as in the present experiments at the LHC.
%
\begin{table}[htb]
\centering
\begin{tabular}{|l|l|l|l|l|l|}
  \hline
  Mode  & $E_e$ (GeV) & $E_A$ (GeV) & $\sqrt{s_{eN}}(eA)$ (GeV) \\
  \hline
  EIC & 5 &  40  & 20 \\
  EIC & 10 &  110  & 47 \\
  EIC & 18 &  110 & 63 \\
  LHeC (á Run1) & 60 & 1380 & 407 \\
  LHeC (á Run2) & 60 & 2760 & 575 \\
  \hline
\end{tabular}
\caption{
     Expected energy ranges accessible by the planned electron-ion colliders at RHIC (EIC)
     \cite{Accardi:2012qut,Aschenauer:2014cki,Aschenauer:2017jsk} and at LHeC \cite{AbelleiraFernandez:2012cc}.
        }
\label{tab:EIC:energy}
\end{table}
%

%
%
%
\subsection{Nonperturbative effects at small $Q^2$}
\label{npt}
%
%
%

\begin{figure}[htb]
  \centering
  \includegraphics[scale=1.38]{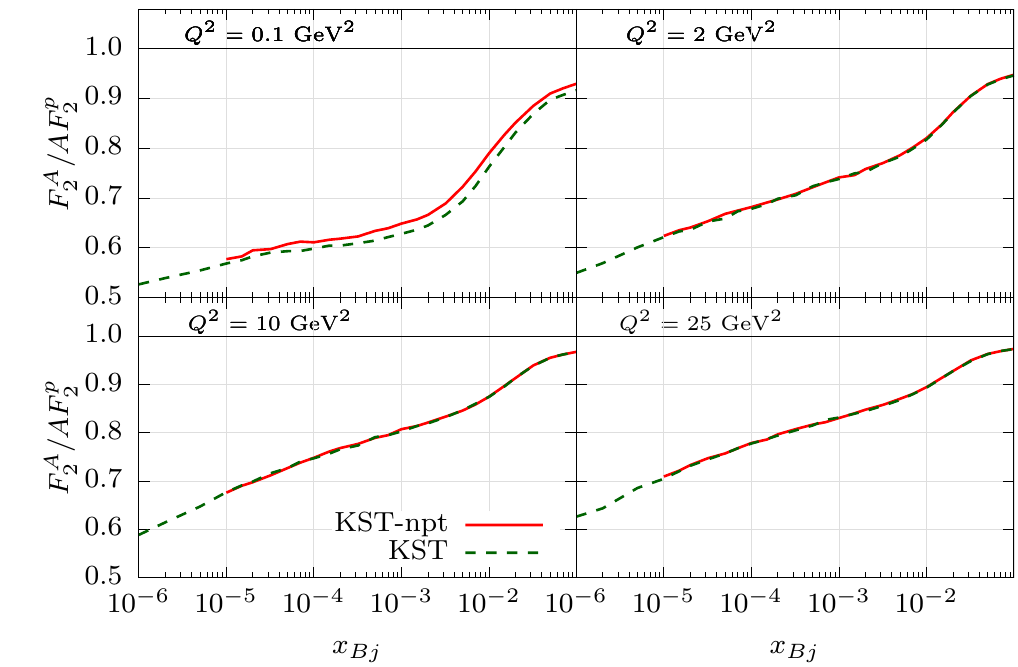}\centering
  \caption{
       The demonstration of the onset of nonperturbative effects in calculations of the quark shadowing (corresponding to the lowest $|q\bar q\ra$ Fock state) for the $Pb$ target as function of Bjorken $x_{Bj}$ at different fixed $Q^2$. The solid and dashed curves correspond to calculations using nonperturbative photon wave functions (see Eqs.~ (\ref{199a}) and (\ref{199b})) and the standard perturbative form (see Eqs.~(\ref{197a}) and (\ref{197b})), respectively.
         }
  \label{fig:Pb-npt}
\end{figure}

As the first step, treating the lowest $|q\bar q\ra$ Fock component, we test the onset of nonperturbative effects adopting the nonperturbative photon wave functions from \cite{Kopeliovich:2008ek} (see also Eqs.~(\ref{199a}) and (\ref{199b})).
Fig.~\ref{fig:Pb-npt} demonstrates that the relative contribution of these effects to nuclear shadowing cannot be neglected only at very small photon virtualities $Q^2 \lsim 2\,\GeV^2$.

%
%

%
%
%
\subsection{Nuclear shadowing from the parton model}
\label{Sec:parton}
%
%
%

In the present paper for more comprehensive study of shadowing effects, besides the color dipole formalism, we express the structure function ratio $F_2^A/(A\,F_2^p)$ also in terms of parton distribution functions
using the following relation,
%
\begin{equation}
\label{parton}
  \frac{F_2^A(x_{Bj},Q^2)}{A\,F_2^p(x_{Bj},Q^2)}
  =
  \frac{
  \sum_f Z_f^2 \left\{R_f^A(x_{Bj},Q^2)\,q_f(x_{Bj},Q^2) +
  R_{\bar f}^A(x_{Bj},Q^2)\, \bar q_{\bar f}(x_{Bj},Q^2) \right\}}
  {
  \sum_f Z_f^2 \left\{q_f(x_{Bj},Q^2) + \bar q_{\bar f}(x_{Bj},Q^2) \right\}}\, ,
\end{equation}
%
where $Z_f$ is the quark charge,
$q_f(x_{Bj},Q^2)$ is the distribution function of a parton $f$ (here we used the CT10 parametrization \cite{Guzzi:2011sv}), the factor $R_{f}^A(x_{Bj},Q^2)$ is the nucleus-to-nucleon ratio of distribution functions for a parton $f$ (parton nuclear modification factor). Here we used the latest EPPS16 parametrization from Ref.~\cite{Eskola:2016oht}. Apart, we also employ the nCTEQ15 parametrization \cite{Kovarik:2015cma} that includes the nuclear parton distribution function in the CTEQ framework.

Comparison of the magnitudes of shadowing using the both color dipole formalism and Eq.~(\ref{parton}) based on the parton model is presented below in Sect.~\ref{Sec:results-EIC}.

%
%
%
\subsection{Predictions for shadowing from the lowest $|q\bar q\ra$ component of the photon}
\label{Sec:q-shad}
%
%
%

%
\begin{figure}[!htb]
  \centering
  \includegraphics[scale=1.4]{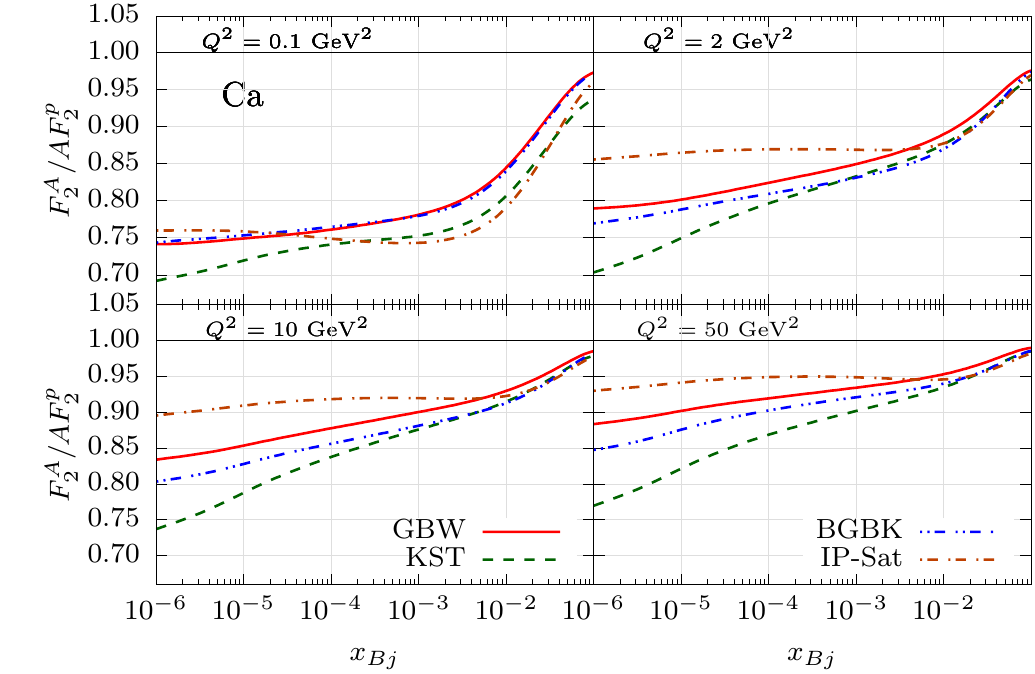}\centering
  \caption{
       Predictions for the shadowing in DIS on $Ca$ target as function of $x_{Bj}$ at several fixed $Q^2$. Calculations correspond to contribution of the lowest $|q\bar q\ra$ Fock component of the photon.
          }
  \label{fig:Ca}
\end{figure}
%
%
\begin{figure}[!htb]
  \centering
  \includegraphics[scale=1.4]{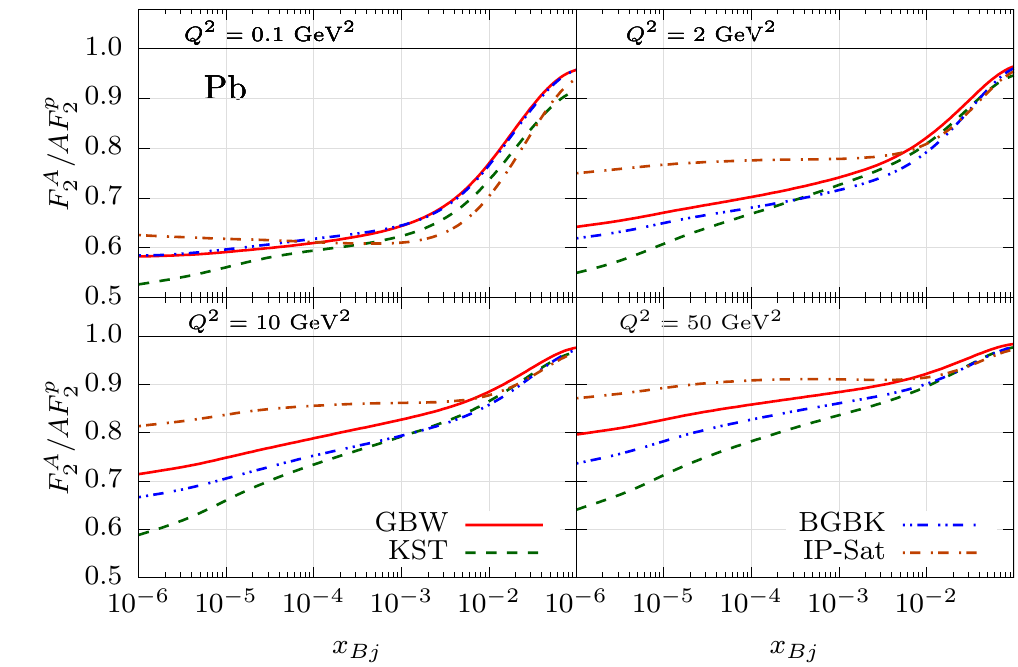}\centering
  \caption{
       The same as Fig.~\ref{fig:Ca} but for the $Pb$ target.
          }
  \label{fig:Pb}
\end{figure}
%

Here, we present the contribution to nuclear shadowing from the lowest
$|q\bar q\ra$ Fock fluctuation of the photon. Calculations have been performed
according to Eq.~(\ref{420}) based on the Green function formalism with
the corresponding exact numerical solution of the evolution equation (see Eq.~(\ref{370})). This allowed to adopt the realistic nuclear density functions
parametrized as is described in Ref.~\cite{DeJager:1987qc}, as well as
realistic parametrizations of the dipole cross section, such as GBW, KST, BGBK and IP-sat
used in our analysis.
The predictions for expected kinematic regions of $x_{Bj}$ and $Q^2$ in experiments at EICs
are presented in
Figs.~\ref{fig:Ca} and \ref{fig:Pb} for the $Ca$ and $Pb$ targets, respectively.

One can see that differences in predictions using various dipole models rises
towards smaller values of $x_{Bj}$. This gives an opportunity to test such
models by the more precise data from EICs.

%
%
%
\subsection{
Predictions for shadowing including higher $|q\bar qG\ra$ Fock component of the photon}
\label{Sec:g-shad}
%
%
%

%
\begin{figure}[!htb]
  \centering
  \includegraphics[scale=1.4]{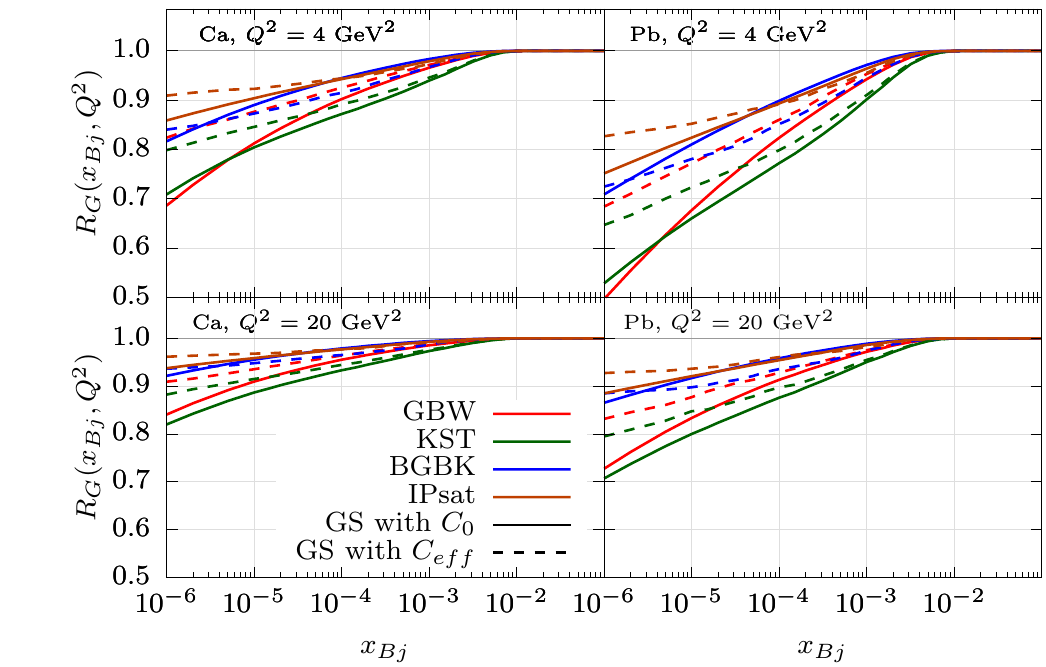}\centering
  \caption{
       Predictions for the gluon shadowing correction from the $q\bar qG$ fluctuation of the photon as function of $\xBj$ for the $Ca$ (left panels) and $Pb$ (right panels) target, respectively. The top and bottom panels include calculations at fixed values of $Q^2=4\,\GeV^2$ and $Q^2=20\,\GeV^2$, respectively. The solid and dashed lines correspond to results using factors $C_0$ and $C_{eff}$, respectively. The magnitude of the gluon shadowing factor is computed using several parametrizations for the dipole cross section as mentioned in the text.
          }
  \label{fig:G}
\end{figure}
%

In this section, we present the results of our calculations of gluon shadowing correction $R_G(x_{Bj},Q^2)$
(see Eq.~(\ref{eq:dipole:gs:RGdef})) corresponding to the $|q\bar q G\ra$ Fock component of the photon containing one gluon. These results are depicted in Fig.~\ref{fig:G}  for the $Ca$ and $Pb$ target at two fixed values of $Q^2 = 4\,\GeV^2$ and $20\,\GeV^2$.
Here, we test several phenomenological parametrizations for $\sigma_{q\bar q}(r)$, such as GBW, KST, BGBK and IP-sat and their impact on the magnitude of the gluon shadowing factor $R_G$.

The Fig.~\ref{fig:G} clearly demonstrates again that uncertainties in predictions of $R_G$, using various dipole models for $\sigma_{q\bar q}(r)$, rise towards small values of Bjorken $\xBj$. Besides, one can see that the magnitude of the gluon shadowing at small $x_{Bj}\lsim 10^{-2}$ is smaller using more realistic factor $C_{eff}$ instead of $C_0$ for all dipole models except for the BGBK parametrization. Such results are consistent also with Fig.~\ref{fig:RatioCeffC0}.

%
%
%
\subsection{Predictions for shadowing expected at planned electron-ion colliders}
\label{Sec:results-EIC}
%
%
%
%
\begin{figure}[!htb]
  \centering
  \includegraphics[scale=1.4]{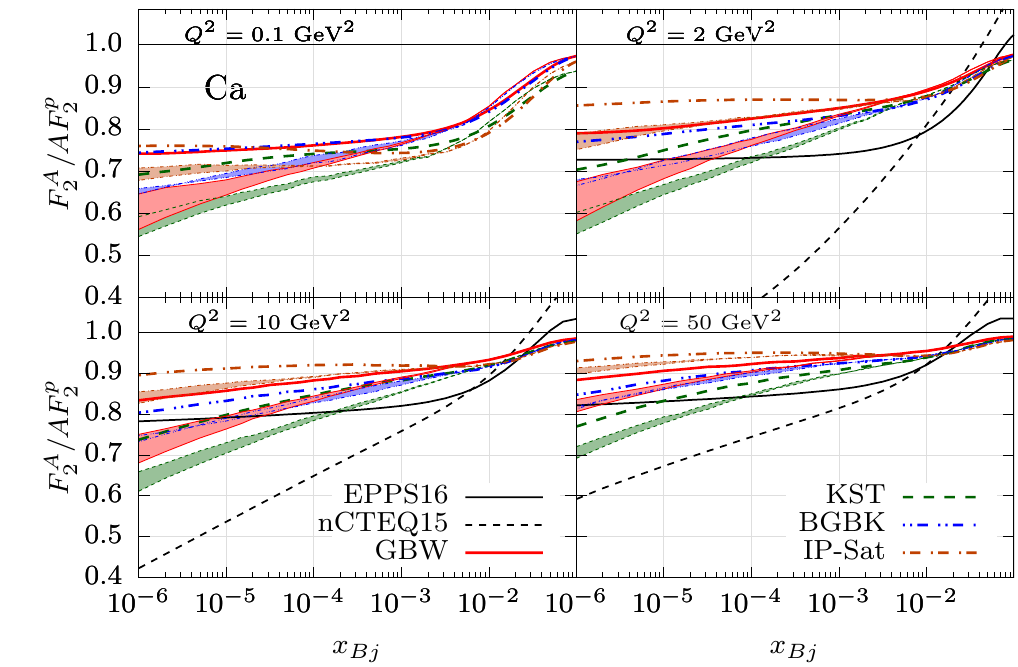}\centering
  \caption{
       Predictions for the overall nuclear shadowing including the contribution from the
       lowest $q\bar q$ photon fluctuations according to Eq.~(\ref{420}), as well as from the higher multigluon Fock states $|q\bar qG...\ra$, effectively included via eikonalization of the gluon suppression factor $R_G(x_{Bj},Q^2)$, according to Eq.~(\ref{eq:dipole:gs:replace}). Calculations have been performed for the $Ca$ target as function of $x_{Bj}$ at different fixed $Q^2$. The boundaries of filled areas correspond to contributions from multi-gluon Fock states calculated with factors $C_0$ and $C_{eff}$. Predictions within the LC dipole formalism for several dipole models (GBW, KST, BGBK, IP-sat) are compared with results based on the parton model, Eq.~(\ref{parton}), using EPPS16 \cite{Eskola:2016oht} and nCTEQ15 \cite{Kovarik:2015cma} parametrizations for the nuclear parton distribution functions.
         }
  \label{fig:GCa-full}
\end{figure}
%
%
\begin{figure}[!htb]
  \centering
  \includegraphics[scale=1.4]{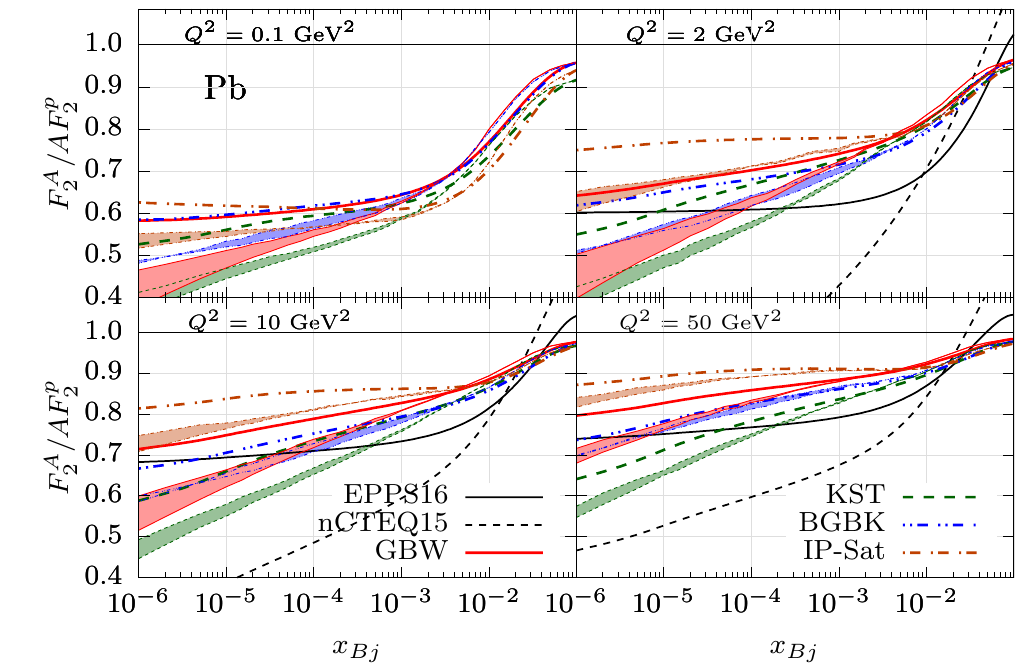}\centering
  \caption{
       The same as Fig.~\ref{fig:GCa-full} but for the $Pb$ target.
          }
  \label{fig:GPb-full}
\end{figure}
%

In this Section, the predictions for the overall nuclear shadowing expected in the kinematic regions accessible by experiments at electron-ion colliders are presented in Figs.~\ref{fig:GCa-full} and \ref{fig:GPb-full} for the calcium and lead target, respectively. Here, we included besides the lowest $|q\bar q\ra$ Fock component of the photon also contributions from higher fluctuations containing gluons relying on Eq.~(\ref{eq:dipole:gs:replace}) \cite{Kopeliovich:2001ee} eikonalizing so the gluon shadowing factor $R_G$ for the $|q\bar qG\ra$ Fock component of the photon.
The filled areas are bounded by calculations of shadowing corrections from the multi-gluon Fock states using factors $C_0$ and $C_{eff}$ as is described above in Sect.~\ref{glue-shad}.

The both Figs.~\ref{fig:GCa-full} and \ref{fig:GPb-full} also show the comparison of our results based on the LC dipole formalism using the Green function technique with the standard results (see Sect.~\ref{Sec:parton}) based on the parton model using the nuclear parton distribution functions. Such a comparison of the shadowing magnitude is performed adopting several dipole models, as well as two parametrizations of nuclear PDFs as is described in the caption of Fig.~\ref{fig:GCa-full}.

Here, we would like to emphasize that differences in predictions for nuclear shadowing related to various dipole models rise towards smaller values of the Bjorken $\xBj$. The sufficiently more precise data on shadowing expected by the future experiments at EIC and LHeC can help to distinguish between various models, mainly in the context to consider their further potential to be employed for shadowing predictions.

Besides the uncertainty coming from various dipole models, we would like to stress the uncertainty affected by the calculation of the gluon shadowing itself which is caused by
different factors $C_0$ and $C_{eff}$ as is discussed  in Sect.~\ref{glue-shad}. However, in comparison with the former, the later uncertainty is much smaller and thus has much a weaker impact on the accuracy of predictions for shadowing as is clearly demonstrated in both Figs.~\ref{fig:GCa-full} and \ref{fig:GPb-full}.

%
%
%
\subsection{Comparison with available data}
\label{Sec:results-data}
%
%
%

%
\begin{figure}[!htb]
  \centering
  \includegraphics[scale=1.55]{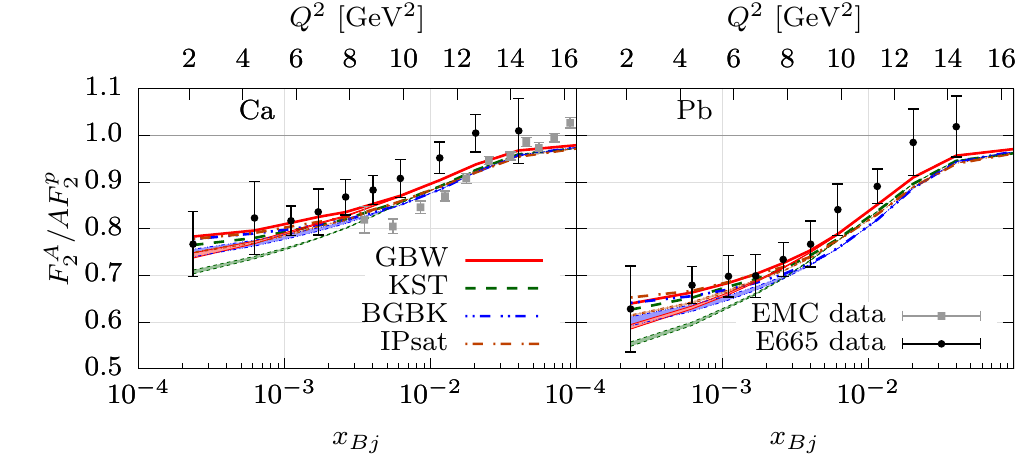}\centering
  \caption{
       Model predictions for the nuclear shadowing vs. data from E665 \cite{Adams:1995is} and EMC \cite{Amaudruz:1995tq} Collaborations for the $Ca$ (left panel) and $Pb$ (right panel) target, respectively. Calculations are performed for several dipole models as indicated in the text.
           }
  \label{fig:data}
\end{figure}
%

The last Fig.~\ref{fig:data} compares our calculations of the shadowing for several dipole models (GBW, KST, BGBK and IP-sat) with available data from the E665 \cite{Adams:1995is} and EMC \cite{Amaudruz:1995tq} Collaborations. The boundaries of shadowed regions correspond to calculations of the gluon shadowing correction using the factors $C_0$ and $C_{eff}$. One can see a reasonable agreement of our predictions with data. However, the error bars at small $x_{Bj}$ are too large for the preference of some dipole model.

%
%
%
\section{Data sets}
\label{datasets}
%
%
%

Numerical data sets that include values for $C_{eff}$ given by Eq.~(\ref{eq:eA:Ceff}) and corresponding gluon shadowing factors $R_G$ are available on Zenodo web-page~ \url{https://zenodo.org/record/3470138} \cite{dataset}.
Here the values of gluon shadowing correction $R_G$ can be computed from Eq.~(\ref{eq:dipole:gs:RGdef}) for $Ca$ and $Pb$ nuclei as a function of $\xBj, Q^2$, and $b$
for various dipole parametrizations of $\sigma_{q\bar q}(r,x)$, such as GBW
\cite{GolecBiernat:1998js,Kowalski:2006hc}, KST \cite{Kopeliovich:1999am},
BGBK \cite{Bartels:2002cj} and IP-sat \cite{Rezaeian:2012ji}.

%
%
%
\section{Conclusions}
\label{conclusions}
%
%
%

In this paper, we present the comprehensive study of the shadowing in deep-inelastic scattering off nuclei in the kinematic regions accessible by the future electron-ion colliders, which will be installed at RHIC and LHC. Model predictions are calculated within the LC color dipole approach based on the rigorous Green function formalism allowing to incorporate naturally the effects of quantum coherence and color transparency.

Calculations of the shadowing, within kinematic regions covered by the future experiments at EICs, allow to include in a sufficient way only contributions from  $|q\bar q\ra$ and $|q\bar qG\ra$ Fock states, safely neglecting the higher multi-gluon fluctuations due to their very large effective mass and, consequently, very weak onset of quantum coherence effects. This enables to perform the proper predictions for the nuclear shadowing without any restrictions for the coherence length. Here, we would like to emphasize that although a very popular Balitsky-Kovchegov equation \cite{Balitsky:1995ub,Kovchegov:1999yj} is able to sum up all Fock components, it does not lead to reliable results since is related to the limit of very long coherence length
when transverse sizes of Fock states are "frozen" during propagation through the nucleus.

In the present paper we compare for the first time the magnitudes of shadowing using various phenomenological models for the dipole cross section, which is inherent in color dipole formalism. We test that our predictions using such models are in a good agreement with available data from the E665 and NMC collaboration. Large error bars especially at small Bjorken $\xBj$ do not allow to exclude any dipole model, used in our analysis, from the potentially reliable description of data in the kinematic regions scanned by the future EICs.

We perform a lot of predictions for the shadowing that can be verified by the corresponding future experiments. This gives a possibility to test various models for the dipole cross section especially at small $\xBj\lsim 10^{-4}$ and, consequently, can shed more light on the onset of low-$\xBj$ saturation phenomena, as well as on effects of nuclear quantum coherence.
More precise data on nuclear shadowing off nuclei from the future experiments can allow to quantify the contribution of the gluon shadowing correction.

Finally, we would like to emphasize that numerical values for the gluon shadowing factor $R_G$ presented in the current paper can also be obtained interactively on Zenodo web-page:
\url{https://zenodo.org/record/3470138} \cite{dataset}.

%
%
%
\section*{Acknowledgements}
%
%
%

J.N. work was partially supported by grants LTC17038 and LTT18002 of the Ministry of  Education,  Youth  and  Sports  of  the  Czech  Republic,  by  the  project  of  the  European Regional Development Fund CZ02.1.01/0.0/0.0/16\_019/0000778 and by the Slovak Funding Agency, Grant 2/0007/18.
The work of M.K. was supported in part by the CONICYT Postdoctorado
N.3180085 (Fondecyt Chile), and by the project Centre of Advanced Applied Sciences with the number: CZ.02.1.01/0.0/0.0/16-019/0000778 (Czech Republic). Project Centre of Advanced Applied Sciences is co-financed by European Union.

\bibliographystyle{unsrt}

\end{document}